\documentclass{LMCS}

\def\doi{9(2:05)2013}
\lmcsheading%
{\doi}
{1--30}
{}
{}
{May.~29, 2011}
{May.~28, 2013}
{}

\usepackage{enumerate,hyperref}

\usepackage{amsfonts}

\begin{document}

\title[Total Representations]{Total Representations}

\author[V.~Selivanov]{Victor Selivanov}  
\address{A.P. Ershov Institute of
Informatics Systems \\
 Siberian Branch of Russian Academy of Sciences\\
Lavrentyev prosp. 6, Novosibirsk, 630090, Russia} 
\email{vseliv@iis.nsk.su}  
\thanks{Supported by DFG-RFBR (Grant
436 RUS 113/1002/01, 09-01-91334), by RFBR Grant 13-01-00015, and
by a Marie Curie International Research Staff Exchange Scheme
Fellowship within the
7th European Community Framework Programme.}   


\ACMCCS{[{\bf Mathematics of computing}]: Continuous
  mathematics---Topology---Point-set topology; Continuous
  mathematics---Continuous functions; [{\bf Theory of computation}]:
  Models of computation---Computability---Turing machines}

\begin{abstract} Almost all representations considered in
computable analysis are partial. We provide arguments in favor of
total representations (by elements of the Baire space). Total
representations make the well known analogy between numberings and
representations closer, unify some terminology, simplify some
technical details, suggest  interesting open questions and new
invariants of topological spaces relevant to computable analysis.
\end{abstract}

\maketitle

\section{Introduction}\label{in}

A {\em numbering} of a set $S$ is a surjection from $\omega$ onto
$S$. Numberings are used to transfer the computability theory on
$\omega$ to many countable structures. A {\em representation} of a
set $S$ is a partial surjection from the Baire space ${\mathcal
N}=\omega^\omega$ onto $S$. Representations are used to transfer the
computability theory on ${\mathcal N}$ to many structures of
continuum cardinality. Numbering theory was mostly systematized by
the Novosibirsk group of researchers in computability theory
\cite{er73a,er75,er77} while representation theory was mostly
systematized by the Hagen group of researchers in computable
analysis (CA) \cite{wei00}.

Although the analogy of representation theory to numbering theory
is well known, there is a striking difference between them:
numberings are in most cases total functions while representations
considered so far are almost  always partial functions (among rare
exceptions are Section 1.1.6 of \cite{ba00} and \cite{br}). Note
that total numberings have a better theory than partial numberings
and are sufficient for many important topics like computable model
theory (although partial numberings also have some advantages, in
particular the corresponding category is known to be cartesian
closed while the category of total numberings is not). One might
expect that total representations are also useful in some
situations.

In this paper we  argue in favor of total representations by
elements of $\mathcal{N}$  which we call just total
representations (TRs). Of course,  similar to the numbering
theory, some properties of partial representations are better than
those of total representations. In particular, the category of
sequential admissibly represented spaces is known to be cartesian
closed \cite{sch02} while the corresponding category of sequential
admissibly totally represented spaces is not (see Section
\ref{admiss}). On the other hand, the class of all sequential
admissibly represented spaces is in a sense too large (it
coincides with the class of sequential $T_0$-spaces which are
quotients of countably based spaces, and only few of such spaces
are really interesting for CA).

From recent results of M. de Brecht \cite{br}  it follows that the
countably based admissibly totally representable spaces coincide
with the so called quasi-Polish spaces (see Section \ref{admiss}
for details). This class of quasi-Polish spaces is a good solution
to the problem from \cite{s08} of finding a natural class of
spaces that includes the Polish spaces and the $\omega$-continuous
domains and has a reasonable descriptive set theory (DST). Note
that while DST for countably based spaces is well understood, this
is not the case for the non-countably based spaces. Since some
non-countably based spaces are quite important for Computation
Theory and CA, the development of some DST for them seems very
desirable. We hope that admissible totally represented spaces
could be of use in such a developments (see Section \ref{admiss}
for additional remarks).

As is well known, representation theory, in contrast to numbering
theory, has a strong topological flavor. In fact, many notions of
CA have two versions: computable and topological, the second being
in a sense a ``limit case'' of the first. The topological version
turns out to be fundamental  for understanding many phenomena
related to the computable version; this is explained by the simple
but important fact that computable functions between topological
spaces are sequentially continuous \cite{sch02}. In this paper we
are mainly concerned with the topological versions, only from time
to time making comments on the computable versions. We have to
warn the reader that TRs are useful mostly for understanding the
topological aspects of CA.

Apart from CA, TRs  are useful also for other fields, in
particular for the hierarchy theory. In Sections \ref{princip} and
\ref{precomp} we show that all levels of the most popular
hierarchies in arbitrary countably-based spaces have principal TRs
(a notion analogous to the corresponding  notion from numbering
theory) which  turn out to be acceptable and precomplete. This
extends or improves some earlier facts from DST
\cite{ke94,mo80,bra05}. Note that hierarchies are also of primary
interest to CA because they provide natural tools to measure the
topological complexity (known also as the discontinuity degrees
\cite{he93,he96}) of non-computable problems in CA.

We also show in Section \ref{princon} that analogs of the main
attractive properties of admissible representations hold for the
so called principal continuous TRs (defined again by analogy to
some notions from numbering theory).  Section \ref{admiss} relates
principal continuous TRs with admissible (partial) representations
which are very popular in CA. In Section \ref{semilat} we  develop
topological analogs of some results on computable numberings
\cite{er77,er06}, this again demonstrates that topological analogs
of results from computability theory are often easier. We include
also some previous remarks on TRs from \cite{s92,s04,s07a}.

We also observe that such important notions  of DST as the
continuous and Borel  reducibilities of equivalence relations on the
Baire space are analogous to similar notions considered in the
context of numbering theory. Moreover, we show in Section
\ref{equiv} that most of the natural substructures of the
structure of continuous degrees of equivalence relations have
undecidable first-order theories. This is done via establishing a
close relation of this reducibility to a version of one of the
Weihrauch's reducibilities \cite{he93,ksz10}.

Thus, the results of this paper  show that sticking to TRs leads
to some natural classes of spaces, makes the analogy between
numberings and representations closer, unifies  terminology,
simplifies some technical details, suggests interesting open
questions and new invariants of topological spaces relevant to CA.
In contrast, total numberings seem to be less important than
partial numberings in the study of effective theory of countable
topological spaces  \cite{sp01}. Note that, though we do present
quite a few apparently new results in the technical sections of
this paper, we also give a reasonable space to discussing the
analogy with numbering theory and to citing known facts which
confirm our claim that TRs deserve a special attention.

In Section \ref{spaces} we mention the spaces relevant to this
paper and discuss  a technical notion of a family of pointclasses.
In Section \ref{hier} we discuss classical hierarchies of DST in
arbitrary spaces. In Section \ref{naming} several reducibility
relations on TRs are introduced and discussed. In Sections
\ref{princip} and \ref{precomp} we introduce the important notion
of a principal TR and show that natural TRs of levels of the
standard hierarchies in the countably based spaces are principal
and precomplete. Section \ref{princon}  shows that the principal
continuous TRs of spaces hold the main attractive properties of
the admissible representations. In Section \ref{admiss} we discuss
admissible TRs  (putting emphasis on the spaces of open sets of
countably based spaces) which turns out to be an important
subclass of admissible partial representations.   Section
\ref{semilat} investigates   semilattices of TRs of open sets in
countably based spaces. Section \ref{categ} presents the category
of TRs which is useful in some contexts. In Section \ref{equiv} we
discuss some reducibility  notions for equivalence relations on
the Baire space.

\section{Spaces and Pointclasses}\label{spaces}

Here we discuss spaces considered in the sequel and a technical
notion of a family of pointclasses that is useful in hierarchy
theory.

We freely  use the standard set-theoretic notation like $|X|$ for
the cardinality of $X$, $X\times Y$ for the cartesian product,
$pr_X(A)=\{x\mid \exists y\in Y(x,y)\in A\}$ for the projection of
$A\subseteq X\times Y$ to $X$, $Y^X$ for the set of functions
$f:X\to Y$, $P(X)$ for the set of all subsets of $X$. For
$A\subseteq X$, $\overline{A}$ denotes the complement $X\setminus
A$ of $A$ in $X$. For $\mathcal{A}\subseteq P(X)$,
$BC(\mathcal{A})$ denotes the Boolean closure of $\mathcal{A}$,
i.e. the set of finite Boolean combinations of sets in
$\mathcal{A}$.

We assume the reader to be familiar with basic notions of
topology. The set of open subsets of a space $X$ is sometimes
denoted $O(X)$. We often abbreviate ``topological space'' to
``space''. A space $X$ is {\em Polish} if it is countably based
and metrizable with a metric $d$ such that $(X,d)$ is a complete
metric space. A space $X$ is {\em quasi-Polish} \cite{br} if it is
countably based and quasi-metrizable with a quasi-metric $d$ such
that $(X,d)$ is a complete quasi-metric space. A {\em
quasi-metric} on $X$ is a function from $X\times X$ to the
nonnegative reals such that $d(x,y)=d(y,x)=0$ iff $x=y$, and
$d(x,y)\leq d(x,z)+d(z,y)$. Since for the quasi-metric spaces
different notions of completeness and of a Cauchy sequence are
considered, the definition of quasi-Polish spaces should be made
more precise (see \cite{br} for additional details).  We skip
these details because we will  in fact use other characterizations
of these spaces given in the sequel.

Let $\omega$ be the space of non-negative integers with the
discrete topology. Of course, the spaces
$\omega\times\omega=\omega^2$, and $\omega\sqcup\omega$ are
homeomorphic to $\omega$, the first homeomorphism is realized by
the Cantor pairing function $\langle x,y\rangle$.

Let ${\mathcal N}=\omega^\omega$ be the set of all infinite
sequences of natural numbers (i.e., of all functions
$\xi:\omega\rightarrow\omega$). Let $\omega^*$ be the set of
finite sequences of elements of $\omega$, including the empty
sequence. For $\sigma\in\omega^*$ and $\xi\in{\mathcal N}$, we
write $\sigma\sqsubseteq \xi$ to denote that $\sigma$ is an
initial segment of the sequence $\xi$. By
$\sigma\xi=\sigma\cdot\xi$ we denote the concatenation of $\sigma$
and $\xi$, and by $\sigma\cdot\mathcal{N}$  the set of all
extensions of $\sigma$ in $\mathcal{N}$. For $x\in\mathcal{N}$, we
can write $x=x(0)x(1)\cdots$ where $x(i)\in\omega$ for each
$i<\omega$. For $x\in\mathcal{N}$ and $n<\omega$, let
$x[n]=x(0)\cdots x(n-1)$ denote the initial segment of $x$ of
length $n$.  Notations in the style of regular expressions like
$0^\omega$, $0^\ast 1$ or $0^m1^n$ have the obvious standard
meaning.

Define the topology on $\mathcal{N}$ by taking arbitrary unions of
sets of the form $\sigma\cdot\mathcal{N}$, where
$\sigma\in\omega^*$, as the open sets. The space $\mathcal{N}$
with this topology known as the {\em Baire space} is of primary
importance for DST and CA. The importance stems from the fact that
many countable objects are coded straightforwardly by elements of
$\mathcal{N}$, and it has very specific topological properties. In
particular, it is zero-dimensional and  the spaces ${\mathcal
N}^2$, ${\mathcal N}^\omega$,   $\omega\times{\mathcal
N}={\mathcal N}\sqcup{\mathcal N}\sqcup\cdots$ are homeomorphic to
${\mathcal N}$. The well known homeomorphisms are given by the
formulas $\langle x,y\rangle(2n)=x(n)$ and $\langle
x,y\rangle(2n+1)=y(n)$, $\langle x_0,x_1,\ldots\rangle(\langle
m,n\rangle)=x_m(n)$,  $(n,x)\mapsto n\cdot x$, respectively.

For any finite alphabet $A$ with at least two symbols, let
$A^\omega$ be the set of $\omega$-words over $A$. This set may be
topologized similar to the Baire space. The resulting space is
known as {\em Cantor space} (more often the last name is applied
to the  space $\mathcal{C}=2^\omega$ of infinite binary sequences).
Although in representation theory the Baire and Cantor spaces are
equivalent as the sets of names, in the study of total
representations Baire space is more suitable. The reason is that
Cantor space is compact, hence all its continuous images are also
compact.

We also need the space  $P\omega$ of subsets of $\omega$ with the
Scott topology on the complete lattice $(P(\omega);\subseteq)$.
The basic open sets of this topology are of the form
$\{A\subseteq\omega\mid F\subseteq A\}$ where $F$ runs through the
finite subsets of $\omega$.

We conclude this section by recalling (in a slightly generalized
form) a technical notion from DST. A {\em pointclass} in $X$ is a
subset of $P(X)$. A {\em family of pointclasses} is a family
$\Gamma=\{\Gamma(X)\}$ indexed by arbitrary spaces such that
$\Gamma(X)\subseteq P(X)$ for any space $X$, and
$f^{-1}(A)\in\Gamma(X)$ for any $A\in\Gamma(Y)$ and any continuous
function $f:X\to Y$. In particular, any pointclass $\Gamma(X)$ in
such a family is downward closed under the Wadge reducibility in
$X$. Recall that $B\subseteq X$ is {\em Wadge reducible} to
$A\subseteq X$ (in symbols $B\leq_WA$) if $B=f^{-1}(A)$ for some
continuous function $f$ on $X$. A basic example of a family of
pointclasses is $O=\{O(X)\}$ where $O(X)$ is the set of open sets
in $X$.  There are also two trivial examples of families $E,F$
where $E(X)=\{\emptyset\}$ and $F(X)=\{X\}$ for any space $X$.

We define some operations on families of pointclasses which are
relevant to hierarchy theory. First, we can use the usual
set-theoretic operations pointwise. E.g., the union $\bigcup_i\Gamma_i$ of
families $\Gamma_0,\Gamma_1,\ldots$ is defined by
$(\bigcup_i\Gamma_i)(X)=\bigcup_i\Gamma_i(X)$.

Secondly, a large class of such operations is induced by the
set-theoretic operations of L.V. Kantorovich and E.M. Livenson
which are now better known under the name ``$\omega$-Boolean
operations''. Relate to any $\mathcal{A}\subseteq P(\omega)$ the
operation $\Gamma\mapsto\Gamma_\mathcal{A}$ on families of
pointclasses as follows:
$\Gamma_\mathcal{A}(X)=\{\mathcal{A}(C_0,C_1,\ldots)\mid
C_0,C_1,\ldots\in\Gamma(X)\}$ where
 $$\mathcal{A}(C_0,C_1,\ldots)= \bigcup_{A\in\mathcal{A}}
 ((\bigcap_{n\in A}C_n)\cap(\bigcap_{n\in \overline{A}}\overline{C}_n)).$$

The operation $\Gamma\mapsto\Gamma_\mathcal{A}$ includes many
useful concrete operations including the operation
$\Gamma\mapsto\Gamma_\sigma$ where $\Gamma_\sigma(X)$ is the set
of all countable unions of sets in $\Gamma(X)$, the operation
$\Gamma\mapsto\Gamma_c$ where $\Gamma_c(X)$ is the set of all
complements of sets in $\Gamma(X)$, and the operation
$\Gamma\mapsto\Gamma_d$ where $\Gamma_d(X)$ is the set of all
differences of sets in $\Gamma(X)$. E.g., the first operation is
obtained from the general scheme if $\mathcal{A}$ is the set of
all non-empty subsets of $P(\omega)$.

Finally, we will need the operation $\Gamma\mapsto\Gamma_p$
defined by $\Gamma_p(X)=\{pr_X(A)\mid A\in\Gamma(\mathcal{N}\times
X)\}$.

We will see that some properties of families are preserved by
these operations. In this section we state this for the following
simple property \cite{ke94}. A family of pointclasses $\Gamma$ is
{\em reasonable} if for any numbering $\nu:\omega\to\Gamma(X)$ its
universal set $\{(n,x)\mid x\in\nu(n)\}$ is in
$\Gamma(\omega\times X)$.  Note that the converse implication
($\{(n,x)\mid x\in\nu(n)\}\in\Gamma(\omega\times X)$  implies that
$\nu(n)\in\Gamma(X)$ for all $n<\omega$) holds for any family
because $x\mapsto(n,x)$ is a continuous function from $X$ to
$\omega\times X$. One easily checks that the families $E,F,O$ are
reasonable.

The next result is straightforward, so the proof is omitted.

 \begin{lem}\label{reason}\hfill
 \begin{enumerate}[\em(1)]
 \item If  $\Gamma$ is a reasonable family of pointclasses then $\Gamma_\sigma$ is
 reasonable.
 \item Let $\Gamma$ be reasonable and $\mathcal{A}\subseteq
 P(\omega)$. Then $\Gamma_\mathcal{A}$ is reasonable.
 \item If $\Gamma$ is reasonable then so is also $\Gamma_p$.
 \end{enumerate}
 \end{lem}

\section{Hierarchies}\label{hier}

Here we briefly discuss some  hierarchies of subsets in arbitrary
spaces which are often of use in DST and CA. First we recall
definition of Borel hierarchy  in arbitrary spaces from
\cite{s04a} (some particular cases were considered in
\cite{sco76,ta79,s82a,s84}). Let $\omega_1$ be the first
non-countable ordinal.

\begin{defi}\label{d-bh}
 Define the sequence
$\{{\bf\Sigma}^0_\alpha(X)\}_{\alpha<\omega_1}$ of pointclasses in
arbitrary space $X$ by induction on $\alpha$ as follows:
${\bf\Sigma}^0_0(X)=\{\emptyset\}$, ${\bf\Sigma}^0_1(X)$ is the
class of open sets in $X$, ${\bf\Sigma}^0_2(X)$  is the class of
countable unions of finite Boolean combinations of open sets, and
${\bf\Sigma}^0_\alpha(X)$ for $\alpha>2$ is the class of countable
unions  of sets in $\bigcup_{\beta<\alpha}{\bf\Pi}^0_\beta(X)$,
where
${\bf\Pi}^0_\beta(X)=\{A\mid\overline{A}\in{\bf\Sigma}^0_\beta(X)\}$.
\end{defi}

The sequence $\{{\bf\Sigma}^0_\alpha(X)\}_{\alpha<\omega_1}$ is
called {\em Borel hierarchy} in $X$. The pointclasses
${\bf\Sigma}^0_\alpha(X)$, ${\bf\Pi}^0_\alpha(X)$ are the {\em
non-selfdual levels} and
${\bf\Delta}^0_\alpha(X)={\bf\Sigma}^0_\alpha(X)\cap{\bf\Pi}^0_\alpha(X)$
are the {\em self-dual levels} of the  hierarchy  (as is usual in
DST, we apply the last terms also to levels of other hierarchies
below).  The pointclass ${\mathbf B}(X)$ of {\em Borel sets} in
$X$ is the union of all levels of the Borel hierarchy. Let us state
the inclusions of levels which are well known for Polish spaces.

\begin{prop}\label{p-bh}
For any space $X$ and for all $\alpha,\beta$ with $\alpha<\beta<\omega_1$,
${\bf\Sigma}^0_\alpha(X)\subseteq{\bf\Delta}^0_\beta(X)$.
\end{prop}

\begin{rem} Definition {\ref{d-bh}} applies to arbitrary
topological space, and Proposition {\ref{p-bh}}
holds true in the full generality. Note that Definition
{\ref{d-bh}} differs from the classical definition for Polish
spaces \cite{ke94} only for the level 2, and that for the case of
Polish spaces our definition of Borel hierarchy is equivalent to
the classical one. The classical definition applied, say, to
$\omega$-continuous domains does not in general have the properties one
expects from a hierarchy. E.g., Proposition {\ref{p-bh}} is true
for our definition  but is in general false for the classical one.
\end{rem}

Note that, in notation of the previous section,  we have
${\bf\Sigma}^0_0=E$, ${\bf\Sigma}^0_1=O$,
${\bf\Sigma}^0_2=(({\bf\Sigma}^0_1)_d)_\sigma\}$ (because
${\bf\Sigma}^0_2(X)$ obviously coincides with the set of countable
unions of differences of open sets in $X$),
${\bf\Sigma}^0_{\alpha+1}=(({\bf\Sigma}^0_\alpha)_c)_\sigma$ for
any countable $\alpha\geq2$, and
${\bf\Sigma}^0_\lambda=(\bigcup_{\alpha<\lambda}{\bf\Sigma}^0_\alpha)_\sigma$
for any limit countable ordinal $\lambda$. Thus, by Lemma \ref{reason} any
fixed non-self-dual level of  Borel hierarchy is a reasonable
family of pointclasses.

For any non-zero ordinal $\theta<\omega_1$, let
$\{{\bf\Sigma}^{-1,\theta}_\alpha\}_{\alpha<\omega_1}$ be the
Hausdorff difference hierarchy over ${\bf\Sigma}^0_\theta$. We
recall the definition. An ordinal $\alpha$ is {\em even}  (resp.
{\em odd}) if $\alpha=\lambda+n$ where $\lambda$ is either zero or
a limit ordinal and $n<\omega$, and the number $n$ is even (resp.,
odd). For an ordinal $\alpha$, let $r(\alpha)=0$ if $\alpha$ is
even and $r(\alpha)=1$, otherwise. For any ordinal $\alpha$,
define the operation $D_\alpha$ sending sequences of sets
$\{A_\beta\}_{\beta<\alpha}$ to sets by
 $$D_\alpha(\{A_\beta\}_{\beta<\alpha})=
\bigcup\{A_\beta\setminus\bigcup_{\gamma<\beta}A_\gamma\mid
\beta<\alpha,\,r(\beta)\not=r(\alpha)\}.$$
 For any ordinal $\alpha<\omega_1$ and  any pointclass ${\mathcal E}$ in $X$,
let $D_\alpha({\mathcal E})$ be the class of all sets
$D_\alpha(\{A_\beta\}_{\beta<\alpha})$, where $A_\beta\in{\mathcal
E}$ for all $\beta<\alpha$. Finally, let
${\bf\Sigma}^{-1,\theta}_\alpha(X)=D_\alpha({\bf\Sigma}^0_\theta(X))$
for any space $X$ and for all $\alpha,\theta<\omega$, $\theta>0$.

It is easy to see that for any $\alpha<\omega_1$  there is
$\mathcal{D}_\alpha\subseteq P(\omega)$ such that
$\mathcal{D}_\alpha({\bf\Sigma}^0_\theta(X))={\bf\Sigma}^{-1,\theta}_\alpha(X)$
for all non-zero $\theta<\omega_1$ and all $X$. Thus, by Lemma \ref{reason}
any fixed non-self-dual level of the difference
hierarchy is a reasonable family of pointclasses. It is well
known and easy to check that
$\bigcup_{\alpha<\omega_1}{\bf\Sigma}^{-1,\theta}_\alpha(X)\subseteq
{\bf\Delta}^0_{\theta+1}(X)$ for all  $0<\theta<\omega_1$ and $X$.

Let $\{{\bf\Sigma}^1_n(X)\}_{1\leq n<\omega}$ be the Luzin's
projective hierarchy in $X$. Using the corresponding operation on
families of pointclasses from the previous section we have
${\bf\Sigma}^1_1(X)=({\bf\Pi}^0_2(X))_p$ and
${\bf\Sigma}^1_{n+1}(X)=({\bf\Pi}^1_n(X))_p$ for any $n\geq1$. The
reason why the definition of the first level is distinct from the
classical definition ${\bf\Sigma}^1_1(X)=({\bf\Pi}^0_1(X))_p$ for
Polish spaces  is again the difference of our definition of
${\bf\Sigma}^0_2$ from the classical one. Again, by Lemma
\ref{reason} any fixed non-self-dual level of the projective
hierarchy is a reasonable family of pointclasses.  It is well
known and easy to check that $\bigcup_{\alpha<\omega_1}{\mathbf
\Sigma}^0_\alpha(X)\subseteq{\bf\Delta}^1_1(X)$.  It is easy to
see that, similar to a known fact for Polish spaces,
${\mathbf\Sigma}^1_1=\mathbf{B}_p$.

For a further reference, we summarize some of the above remarks.

 \begin{lem}\label{hierop}\hfill
 \begin{enumerate}[\em(1)]
 \item In notation of the previous section,
${\bf\Sigma}^0_0=E$, ${\bf\Sigma}^0_1=O$,
${\bf\Sigma}^0_2=(({\bf\Sigma}^0_1)_d)_\sigma$,
${\bf\Sigma}^0_{\alpha+1}=(({\bf\Sigma}^0_\alpha)_c)_\sigma$ for
any countable $\alpha\geq2$, and
${\bf\Sigma}^0_\lambda=(\bigcup_{\alpha<\lambda}{\bf\Sigma}^0_\alpha)_\sigma$ for any
limit countable ordinal $\lambda$.
 \item For any $\alpha<\omega_1$  there is
$\mathcal{D}_\alpha\subseteq P(\omega)$ such that
$\mathcal{D}_\alpha({\bf\Sigma}^0_\theta)={\bf\Sigma}^{-1,\theta}_\alpha$
for all non-zero $\theta<\omega_1$.
 \item We have ${\bf\Sigma}^1_1=({\bf\Pi}^0_2)_p$ and
${\bf\Sigma}^1_{n+1}=({\bf\Pi}^1_n)_p$ for any $n\geq1$.
 \item Any non-self-dual level of any of the three hierarchies is
a reasonable family of pointclasses.
 \end{enumerate}
\end{lem}

\noindent For levels of the difference and projective hierarchies we
have the natural inclusions similar to those in Proposition
\ref{p-bh}.  Note that for Polish spaces the class ${\bf\Sigma}^1_{1}$
of analytic sets has several nice equivalent characterizations (in
particular, as the class of continuous images of Polish spaces or as
the class of sets obtained from the closed sets by applying the Suslin
A-operation). In Lemma 56 of \cite{br} it was observed that the
characterization in terms of continuous images extends to the
quasi-Polish spaces. In Section \ref{admiss} we will see that this
characterization fails in general for non-countably based admissibly
totally represented spaces.

Next we establish  important structural properties of
$\mathbf{\Sigma}$-levels of the Borel hierarchy which are well
known for Polish spaces \cite{ke94}. This result  demonstrates
that our extension of the classical definition to arbitrary spaces
is natural.

Let $\Gamma$ be a family of pointclasses. A pointclass $\Gamma(X)$
has the {\em $\omega$-reduction property} if  for each countable
sequence $A_0, A_1,\ldots$ in $\Gamma(X)$ there is a countable
sequence $D_0, D_1,\ldots$ in $\Gamma(X)$ such that $D_i\subseteq
A_i$, $D_i\cap D_j=\emptyset$ for all $i\not=j$ and
$\bigcup_{i<\omega}D_i=\bigcup_{i<\omega}A_i$. A pointclass
$\Gamma(X)$  has the {\em  $\omega$-uniformization property} if
for any $A\in\Gamma(\omega\times X)$ there is
$D\in\Gamma(\omega\times X)$ such that $D\subseteq A$,
$pr_X(D)=pr_X(A)$, and for any $x\in X$ there is at most one
$n\in\omega$ with $(n,x)\in D$;  we say that such set $D$ {\em
uniformizes} $A$. Just as in \cite{ke94} one can check that if
$\Gamma$ is reasonable then $\Gamma(X)$ has the
$\omega$-uniformization property iff it has the $\omega$-reduction
property.

\begin{thm}\label{uniform}
For any space $X$ and any $2\leq\alpha<\omega_1$,
${\bf\Sigma}^0_\alpha(X)$ has the $\omega$-reduction and
$\omega$-uniformization properties. If $X$ is zero-dimensional,
the same holds for the class ${\bf\Sigma}^0_1(X)$ of open sets.
\end{thm}

\proof By item 4 of Lemma \ref{hierop} and remarks before
the formulation, it suffices to establish the
$\omega$-uniformization property. We consider only the second
level. For $\alpha>2$ and $\alpha=1$ the proof is almost the same.

Let $A\in{\bf\Sigma}^0_2(\omega\times X)$, then
$A=\bigcup_n(B_n\setminus C_n)$ for some
$B_n,C_n\in{\bf\Sigma}^0_1(\omega\times X)$. Then $x\in pr_X{A}$
iff $\exists n(x\in A(n))$ iff $\exists n,m(x\in B_m(n)\setminus
C_m(n))$ where $A(n)=\{x\mid (n,x)\in A\}$. Let
 $$E_{m,n}=\{x\mid x\in B_m(n)\setminus
C_m(n)\wedge\neg\exists\langle m_1,n_1\rangle<\langle m,n\rangle
(x\in B_{m_1}(n_1)\setminus C_{m_1}(n_1))\}.$$
 Then $E_{m,n}\cap E_{m_1,n_1}=\emptyset$ for any distinct $\langle m,n\rangle$
and $\langle m_1,n_1\rangle$, and $E_{m,n}$ is a finite Boolean
combination of open sets. Therefore, $D_n=\bigcup_mE_{m,n}$ is in
${\bf\Sigma}^0_2(X)$ for any $n$, hence $D=\{(n,x)\mid x\in D_n\}$
is in ${\bf\Sigma}^0_2(\omega\times X)$. The set $D$ uniformizes
$A$.
\qed

For arbitrary spaces, not much can be said about more interesting
properties of the introduced hierarchies like the non-collapse
property saying that any ${\bf\Sigma}$-level is distinct from the
corresponding ${\bf\Pi}$-level. We come back to such non-trivial
questions in Section \ref{princip}.

In \cite{br} the following important characterization of
quasi-Polish spaces in terms of Borel hierarchy was obtained.

\begin{prop}\label{pi2}
A space is quasi-Polish iff it is homeomorphic to a ${\bf
\Pi}^0_2$-subset of $P\omega$ with the induced topology.
\end{prop}

The computable versions of the introduced hierarchies are defined
in a straightforward way \cite{s06}  but their non-trivial
properties (like the effective Hausdorff-Kuratowski theorem) seem
to be relatively well understood only for the spaces $\omega$,
$\mathcal{N}$ and $\mathcal{C}$. To my knowledge, the problem of
finding a broad enough class of effective spaces with good
effective DST is  open.

\section{Representations and Reducibilities}\label{naming}

In this section we introduce and briefly discuss some reducibility
notions which serve as tools for measuring the topological
complexity of problems in DST and CA.

By a {\em total representation} (TR) we mean any function $\nu$
with $dom (\nu)=\mathcal{N}$. By a {\em total representation of a
given set $S$} we mean a TR $\nu$ with $rng (\nu)=S$. There are
several natural reducibility notions for TRs the most basic of
which is the following. A TR $\mu$ is {\em reducible} to a TR
$\nu$ (in symbols $\mu\leq\nu$) if $\mu=\nu\circ f$ for some
continuous function $f$ on  $\mathcal{N}$.  A TR $\mu$ is {\em
equivalent} to $\nu$ (in symbols $\mu\equiv\nu$), if $\mu\leq\nu$
and $\nu\leq\mu$.

For any set $S$, we may form the preorder $(S^\mathcal{N};\leq)$
which generalizes the preorder formed by the classical Wadge
reducibility on subsets of $\mathcal{N}$. Indeed, for
$S=2=\{0,1\}$ the structures $(P(\mathcal{N});\leq_W)$ and
$(S^\mathcal{N};\leq)$ are isomorphic: $A\leq_W B$ iff $c_A\leq
c_B$ where $c_A:\mathcal{N}\to 2$ is the characteristic function
of a set $A\subseteq\mathcal{N}$. Note that the structure
$(S^\mathcal{N};\leq)$ (more precisely, its quotient-structure) is
an upper semilattice with the join operation induced by the binary
operation $\oplus$ on $S^\mathcal{N}$ defined by:
$(\mu\oplus\nu)(2n\cdot x)=\mu(x)$, and $(\mu\oplus\nu)((2n+1)\cdot
x)=\nu(x)$. In fact,  this semilattice is a {\em
$\sigma$-semilattice}  \cite{s07a}, i.e. any countable set of
elements has a supremum; the supremum operation is induced by the
operation $(\bigsqcup_n\nu_n)(n\cdot x)=\nu_n(x)$ on sequences
$\{\nu_n\}$ of TRs.

We will also need the  unary operations $p_s$ ($s\in S$) on
$S^\mathcal{N}$ introduced in \cite{s04} defined by:
$[p_s(\nu)](a)=s$, if $a\not\in 0^\ast 1$, and
$[p_s(\nu)](a)=\nu(b)$ otherwise, where $a=0^n1b$ for some
$n<\omega$. We need the following properties of the introduced
operations established in \cite{s04}.  The properties of these
operations are similar to the properties of completion operations
in the theory of complete numberings \cite{s82,s04}.

\begin{prop}\label{sdc}
The quotient-structure of $(S^\mathcal{N};\leq,\oplus,p_s)$ is a
semilattice with discrete closures, i.e.:  $\oplus$ is a supremum
operation for $\leq$; $\nu\leq p_s(\nu)$, $\mu\leq\nu\rightarrow
p_s(\mu)\leq p_s(\nu)$, and $p_s(p_s(\nu))\leq p_s(\nu)$;
 $p_s(\mu)\leq p_u(\nu)\wedge s\not=u\rightarrow p_s(\mu)\leq\nu$;
 $p_s(\mu)\leq\nu\oplus\xi\rightarrow p_s(\mu)\leq\nu\vee p_s(\mu)\leq\xi$.
Moreover, if $f:S\rightarrow T$ then
$f\circ(\mu\oplus\nu)=(f\circ\mu)\oplus(f\circ\nu)$ and $f\circ
p_s(\nu)=p_{f(s)}(f\circ\nu)$.
\end{prop}

The structure of Wadge degrees (i.e., the quotient-structure of
$(P(\mathcal{N});\leq_W)$) is fairly well understood and turns our
to be rather simple.  In particular,
$(\mathbf{\Delta}^1_1(\mathcal{N});\leq_W)$ is almost well ordered
\cite{wad84}, i.e. it has no infinite descending chain and for any
$A,B\in\mathbf{\Delta}^1_1(\mathcal{N})$ we have $A\leq_WB$ or
$\overline{B}\leq_WA$. Beyond the Borel sets, the structure of
Wadge degrees depends on the set-theoretic axioms but under some
of these axioms the whole structure remains almost well ordered.
This structure includes and refines the structure of levels (more
precisely, of the Wadge complete sets in these levels) of the
hierarchies from the previous section (taken for the Baire space).
It may serve as a nice tool to measure the topological complexity
of many problems of interest in DST and CA.

In particular, we will see below that some natural classes of
TRs and of spaces may be defined through the {\em kernel}
$E_\nu=\{\langle a,b\rangle |\nu(a)=\nu(b)\}$ of a TR $\nu$.
The kernel is a subset of $\mathcal{N}$ that codes the
corresponding equivalence relation on $\mathcal{N}$. Clearly,
$\mu\leq\nu$ implies $E_\mu\leq_WE_\nu$ but not vice versa.  Note
that the kernel relation of a given numbering is rather important
in numbering theory.

Already for $3\leq k<\omega$ the structures $(k^\mathcal{N};\leq)$
of $k$-partitions of $\mathcal{N}$ (i.e.,  of TRs of subsets of
$k$) become much more complicated. Nevertheless, some important
information on these structures is already available. For any
$\mathcal{A}\subseteq P(\mathcal{N})$, let $\mathcal{A}_k$ denote
the set of $k$-partitions $\nu\in k^\mathcal{N}$ such that
$\nu^{-1}(i)\in\mathcal{A}$ for each $i<k$. In \cite{ems87} it was
shown that the structure
$((\mathbf{\Delta}^1_1(\mathcal{N}))_k;\leq)$ is a well preorder,
i.e. it has neither infinite descending chains nor infinite
antichains. In \cite{he93,s07a} the quotient-structures of
$((BC(\mathbf{\Sigma}^0_1))_k;\leq)$ and
$((\mathbf{\Delta}^0_2)_k;\leq)$ over $\mathcal{N}$ were
characterized in terms of a natural preorder $\leq_h$ on the
finite and countable well-founded $k$-labeled forests,
respectively. These characterizations clarified the corresponding
structures considerably and led to deep definability theories for
both structures in \cite{ks07,ks09,ksz09}. These results show
that, similar to the structure of Wadge degrees, the structures of
degrees of $k$-partitions may serve as tools to measure the
topological complexity of natural problems. For wider classes of
$k$-partitions like $((\mathbf{\Delta}^0_3)_k;\leq)$, the
corresponding characterizations are not yet known. An impression
on how they can look can be obtained in \cite{s07a,s11} where the
structure of Wadge degrees of regular (in the sense of automata
theory) $k$-partitions of the Cantor space is characterized.

For a further reference we recall some details of the  results in
\cite{he93,s07a}. A poset $(P;\leq)$ will be often shorter denoted
just by $P$. Any subset of $P$ may be considered as a poset with
the induced partial ordering. In particular, this applies to the
``cones" $\uparrow{x}=\{y\in P\mid x\leq y\}$ and
$\downarrow{x}=\{y\in P\mid y\leq x\}$ defined by any $x\in P$. By
{\em a forest} we mean a finite poset  in which every lower cone
$\downarrow{x}$ is a chain. {\em A tree} is a forest having a
smallest element (called {\em the root} of the tree). Note that
any  forest is uniquely representable as a disjoint union of
trees, the roots of the trees being  the minimal elements of the
forest. Let ${\mathcal P}$ (resp. ${\mathcal F}$)  denote the set
of all finite posets (resp. forests) with $P\subseteq\omega$.

We relate to any $F\in{\mathcal F}$ the TR  $\xi_F\in F^\mathcal{N}$
by induction on $|F|$ as follows: if $F=\{r\}$ then $\xi_F=\lambda
x.r$; if $F$ is a non-singleton tree with a root $r$ then
$\xi_F=p_r(\xi_{F\setminus\{r\}})$; if $F=T_0\cup\cdots\cup T_n$ is
a disjoint union of trees where $n>0$ then
$\xi_F=\xi_{T_0}\oplus\cdots\oplus\xi_{T_n}$.  It is easy to see
that $\xi_F$ is an admissible TR of $F$ with respect to the Scott
topology on the forest $F$.

A {\em $k$-labeled poset} (or just a $k$-poset) is an object
$(P;\leq,c)$ consisting of a finite poset $(P;\leq)$ and a labeling
$c:P\rightarrow k$. Sometimes we simplify notation of a $k$-poset to
$(P,c)$ or even to $P$.  A {\em  morphism}
$f:(P;\leq,c)\rightarrow(P^\prime;\leq^\prime,c^\prime)$ between
$k$-posets is a monotone function
$f:(P;\leq)\rightarrow(P^\prime;\leq^\prime)$ respecting the
labelings, i.e. satisfying $c=c^\prime\circ f$.
Let ${\mathcal P}_k$ (resp. ${\mathcal F}_k$)  be the set of all finite
$k$-posets (resp. $k$-forests) $(P;\leq,c)$ with $P\subset\omega$.

Define  a preorder $\leq_h$ on ${\mathcal P}_k$ as follows:
$(P,c)\leq(P^\prime,c^\prime)$, if there is a morphism from $(P,c)$
to $(P^\prime,c^\prime)$. By $\equiv_h$ we denote the {\em
$h$-equivalence relation} on ${\mathcal P}_k$ induced by $\leq_h$.
For $T_0,\ldots,T_n\in{\mathcal F}_k$, let $F=T_0\sqcup\cdots\sqcup
T_n$ be their  disjoint union, then $F\in{\mathcal F}_k$. For
$F\in{\mathcal F}_k$ and  $i<k$, let $p_i(F)$ be the $k$-tree
obtained from $F$ by joining a new bottom element  (from $\omega$)
and assigning the label $i$ to the bottom element. It is clear that
any $k$-forest is $h$-equivalent to a term of signature
$\{\sqcup,p_0,\ldots,p_{k-1},0,\ldots,k-1\}$ without free variables
(the constant symbol $i$ in the signature is interpreted as the
singleton tree carrying the label $i$).

It is known \cite{he93,s04} that the  quotient-structure of
$({\mathcal F}_k;\leq_h)$, together with a new bottom element, is
a distributive lattice any principal ideal of which is finite. The
following assertion from  \cite{s07a} (in which $\sqcup$ is the
binary disjoint union operation) is a version of a much earlier result in
\cite{he93}.

\begin{prop}\label{kfor}
The quotient-structures of the structures $({\mathcal
F}_k;\leq_h,\sqcup,p_0,\ldots,p_{k-1})$  and
$(BC(\mathbf{\Sigma}^0_1(\mathcal{N})))_k;\leq,\oplus,p_0,\ldots,p_{k-1})$
are isomorphic. An isomorphism is induced by the function
$(F;c)\mapsto c\cdot\xi_F$.
\end{prop}

Let us mention some other interesting reducibilities on TRs. A
straightforward generalization of $\leq$ is the reducibility by
functions in $F$ where $F$ is an arbitrary class of functions on
$\mathcal{N}$ closed under composition and containing the identity
function. In particular, let $\leq_{\mathbf{\Delta}^0_\alpha}$
(resp. $\leq_{\mathbf{\Delta}^1_1}$) be the reducibility by
functions $f$ on $\mathcal{N}$ such that
$f^{-1}(A)\in\mathbf{\Delta}^0_\alpha$ for any
$A\in\mathbf{\Delta}^0_\alpha$ (rep.
$f^{-1}(A)\in\mathbf{\Delta}^1_1$ for any
$A\in\mathbf{\Delta}^1_1$). Note that $\leq_{\mathbf{\Delta}^0_1}$
coincides with $\leq$. Some deep facts on the corresponding degree
structures are known, in particular for any countable ordinal
$\alpha>1$ the quotient-structures of
$(\mathbf{\Delta}^1_1(\mathcal{N});\leq_{\mathbf{\Delta}^0_\alpha})$
and $(\mathbf{\Delta}^1_1(\mathcal{N});\leq_W)$ are isomorphic
\cite{an06}. For recent results on similar reducibilities on
arbitrary quasi-Polish spaces see \cite{mss12}

In \cite{wei92,he93,wei00}  some notions of reducibility for functions
on  spaces were introduced which turned out useful for
understanding the non-computability and non-continuity of
interesting decision problems in computable analysis
\cite{he96,bg11a}  and constructive mathematics \cite{bg11}. In
particular, the following notions of reducibilities between
functions $f:X\rightarrow Z$, $g:Y\rightarrow Z$ on topological spaces were
introduced: $f\leq_1g$ (resp.  $f\leq_2g$) iff $f=F\circ g\circ H$
for some continuous functions $H:X\rightarrow Y$ and
$F:Z\rightarrow Z$, (resp. $f(x)=F(x,gH(x))$ for some continuous
functions $H:X\rightarrow Y$ and $F:X\times Z\rightarrow Z$).

Deep results are known for the particular case of these  relations
where $X=Y=\mathcal{N}$ and $Z=k=\{0,\ldots,k-1\}$ is a discrete
space with $k<\omega$ points. In this way we obtain preorders
$(k^\mathcal{N};\leq_1)$ and $(k^\mathcal{N};\leq_2)$. In
\cite{he93} the quotient-structures of
$((BC(\mathbf{\Sigma}^0_1))_k;\leq_1)$ and
$((BC(\mathbf{\Sigma}^0_1))_k;\leq_2)$ were characterized in terms
of  natural preorders on the finite  $k$-labeled forests similar
to the $h$-preorder. These characterizations  led to the proof of
undecidability of first order theories of both quotient-structures
in \cite{ksz10}, for each $k\geq 3$.

In computability theory, numbering theory and CA,  effective
versions of $\leq$ (the reducibility by computable functions on
$\omega$ and $\mathcal{N}$) and of the other reducibilities
mentioned above are extensively studied. Since the corresponding
degree structures become extremely complicated, they cannot serve
as tools for measuring  the computational complexity (in
particular, the degree structures are not well-founded, hence it
is not possible to assign an ordinal to an arbitrary degree). For
this purpose people usually prefer to use complete sets in
suitable effective hierarchies like those discussed in the
previous section. Another way to ``improve'' the algebraic
structure of, say, Weihrauch degrees is to extend the Weihrauch
reducibility to multi-valued functions
\cite{wei92,wei00,bg11,bg11a}. In this way one obtains
algebraically more regular degree structures which are applicable
to the complexity of many interesting problems related to
Constructive Analysis.

\section{Principal Total Representations of Pointclasses}\label{princip}

An important observation in numbering theory is that principal
numberings (i.e., numberings  which are the largest, w.r.t. the
reducibility relation, elements in natural classes of numberings)
are often interesting and have nice properties. Good examples of
principal numberings are the standard computable numberings of
computable partial functions and of computably enumerable sets.

This also applies to DST and CA where principal TRs appear quite
naturally, as we show here and in the sequel. In particular, the
TRs from the following theorem play a crucial role in proving the
non-collapse property of the classical hierarchies from Section
\ref{hier}. Note that some relevant properties of representations
were considered earlier in the context of DST and CA (see e.g.
\cite{ke94,mo80,bra05}).

Let $\Gamma$ be a family of pointclasses. A TR
$\nu:\mathcal{N}\to\Gamma(X)$ is a {\em $\Gamma$-TR} if its
{\em universal set} $U_\nu=\{(a,x)\mid x\in\nu(a)\}$ is in
$\Gamma(\mathcal{N}\times X)$, and $\nu$ is a {\em principal
$\Gamma$-TR} if it is a $\Gamma$-TR and any
$\Gamma$-TR is reducible to $\nu$.  Note that if
$\nu:\mathcal{N}\to\Gamma(X)$ is principal then it is a surjection
and that $\Gamma(X)$ has at most one principal TR, up to
equivalence. Note also that $\nu\mapsto U_\nu$ is a bijection
between the $\Gamma$-TRs $\nu:\mathcal{N}\to\Gamma(X)$ and the
sets in $\Gamma(\mathcal{N}\times X)$ because any
$A\in\Gamma(\mathcal{N}\times X)$ may be considered as the
universal set of the TR $a\mapsto A(a)=\{x\mid (a,x)\in A\}$.
We show that the introduced notions are in a sense preserved by
the operations on families in Section \ref{spaces}.

 \begin{lem}\label{presprin}\hfill
 \begin{enumerate}[\em(1)]
 \item Let $\mathcal{A}\subseteq P(\omega)$ and let $\Gamma$ be a family of
 pointclasses. If $\Gamma(X)$ has a principal $\Gamma$-TR then
$\Gamma_\mathcal{A}(X)$ has a principal
$\Gamma_\mathcal{A}$-TR.
 \item If $\Gamma$ is a family of
pointclasses and $\Gamma(\mathcal{N}\times X)$ has a principal
$\Gamma$-TR then $\Gamma_p(X)$ has a principal
$\Gamma_p$-TR.
 \item If $\{\Gamma_n\}$ is a sequence of families of
pointclasses and $\Gamma_n(X)$ has a principal $\Gamma_n$-TR
for each $n<\omega$  then $(\bigcup_n\Gamma_n)_\sigma(X)$ has a
principal $(\bigcup_n\Gamma_n)_\sigma$-TR.
 \end{enumerate}
\end{lem}

\proof We only define the corresponding TRs, it is
straightforward to verify that they are indeed principal.

1. Let $\nu$ be a principal $\Gamma$-TR of $\Gamma(X)$. Define
the principal $\Gamma_\mathcal{A}$-TR $\nu_\mathcal{A}$ of
$\Gamma_\mathcal{A}(X)$ as follows: $\nu_\mathcal{A}\langle
a_0,a_1,\ldots\rangle=\mathcal{A}(\nu(a_0),\nu(a_1),\ldots)$.

2. Let $\nu$ be a principal $\Gamma$-TR of
$\Gamma(\mathcal{N}\times X)$. Define the principal
$\Gamma_p$-TR $\nu_p$ of $\Gamma_p(X)$ as follows:
$\nu_p(a)=pr_X(\nu(a))$.

3. Let $\nu_n$ be a principal $\Gamma_n$-TR of $\Gamma_n(X)$,
for each $n<\omega$. Define the principal
$(\bigcup_n\Gamma_n)_\sigma$-TR $\nu$  of
$(\bigcup_n\Gamma_n)_\sigma(X)$ as follows: $\nu\langle n_0\cdot
a_0,n_1\cdot
a_1,\ldots\rangle=\nu_{n_0}(a_0)\cup\nu_{n_1}(a_1)\cup\cdots)$.
\qed

The following main result of this section shows that all
non-self-dual levels of the hierarchies from Section \ref{hier}
have principal TRs in any countably based space.  Particular cases
of these results for Polish spaces where known from the early days
of DST \cite{ke94} (see also \cite{s92} for computable versions
and \cite{bra05} for a study of representations of finite levels
of the Borel hierarchy). Our results extend them to a wider class
though the proof remains elementary.

\begin{thm}\label{prin}
Let $X$ be a countably based space and let $\Gamma$ be an
arbitrary non-self-dual level of a hierarchy from Section
\ref{hier}. Then $\Gamma(X)$ has a principal $\Gamma$-TR.
\end{thm}

\proof We consider first the level
$\mathbf{\Sigma}^0_1$. Let $B_0,B_1,\ldots$ be a base in $X$
containing the empty set, say $B_0=\emptyset$. We define the
TR $\pi$  of $\mathbf\Sigma^0_1$  by
$\pi(a)=\bigcup_nB_{a(n)}$. First we have to show that $U_\pi$ is
open in $\mathcal{N}\times X$. For each $(a,x)\in U_\pi$ it
suffices  to find $V\in\mathbf\Sigma^0_1(\mathcal{N}\times X)$
with $(a,x)\in V\subseteq U_\pi$. Since $x\in\pi(a)$, $x\in B_{a(n)}$
for some $n<\omega$. Then $x\in\pi(b)$ for each $b\supseteq a[n+1]$.
Hence we can take $V=(a[n+1]\cdot\mathcal{N})\times B_{a(n)}$.

It remains to show that TR $\pi$ is a largest element in the
corresponding class. Let
$\nu:\mathcal{N}\to\mathbf\Sigma^0_1(X)$ be a $\mathbf\Sigma^0_1$-TR,
so $U_\nu$ is open in $\mathcal{N}\times X$. Define
$f:\mathcal{N}\to\mathcal{N}$ as follows: $f(a)\langle
i,j\rangle=j$ if $B_j\subseteq\bigcap\nu(a[i]\cdot\mathcal{N})$,
and $f(a)\langle i,j\rangle=0$ otherwise. Clearly, $f$ is
continuous (even Lipschitz). It remains to show that $f$ reduces
$\nu$ to $\pi$, i.e. $\nu(a)=\bigcup\{B_j\mid\exists
i(B_j\subseteq\bigcap\nu(a[i]\cdot\mathcal{N}))\}$. Indeed, the
inclusion $\supseteq$ is obvious. Conversely, let $x\in\nu(a)$, so
$(a,x)\in U_\nu$. Since $U_\nu$ is open,
$(a,x)\in(a[i]\cdot\mathcal{N})\times B_j\subseteq U_\nu$ for some
$i,j<\omega$. Then $x\in B_j$ and $y\in\nu(b)$ for all
$b\sqsupseteq a[i]$, $y\in B_j$. Then
$B_j\subseteq\bigcap\nu(a[i]\cdot\mathcal{N})$ hence
$x\in\bigcup\{B_j\mid\exists
i(B_j\subseteq\bigcap\nu(a[i]\cdot\mathcal{N}))\}$.

For the other levels the assertion follows from Lemmas
\ref{hierop} and \ref{presprin}.
\qed

\begin{rem} The TR $\pi$ from  the last proof has many other
interesting properties. In particular, we will see in Section
\ref{admiss} that it is admissible w.r.t. some natural topologies
on $\mathbf{\Sigma}^0_1(X)$.
\end{rem}

\begin{cor}\label{prin1}
Let $X$ be a countably based space and let $\Gamma$ be an
arbitrary non-self-dual level of a
hierarchy from Section \ref{hier}. Then there is a
Wadge-complete set in $\Gamma(\mathcal{N}\times X)$.
\end{cor}

\proof Let $\nu$ be a principal TR of $\Gamma(X)$. We
claim that $U_\nu$ is Wadge-complete in $\Gamma(\mathcal{N}\times
X)$. Let $A\in\Gamma(\mathcal{N}\times X)$, we have to show
$A\leq_WU_\nu$. Since $A$ is the universal set of the TR
$a\mapsto A(a)$, this TR is a $\Gamma$-TR, hence
$A(a)=\nu f(a)$ for some continuous function $f$ on $\mathcal{N}$.
Then $A\leq_WU_\nu$ via the continuous function $\langle
a,x\rangle\mapsto\langle f(a),x\rangle$.
\qed

Using diagonalization,  one immediately derives from Proposition
\ref{prin} the non-collapse property for all three hierarchies in
the Baire space. The non-collapse property is known  to hold in
any uncountable Polish space \cite{ke94}. Recently this was extended
\cite{br} (at least for the Borel and Luzin hierarchies) to any
uncountable quasi-Polish space.

For Polish spaces $X$ the following important relationships between
the introduced hierarchies are known \cite{ke94}:
$\bigcup_{\alpha<\omega_1}{\bf \Sigma}^0_\alpha(X)={\bf\Delta}^1_1(X)$
(Suslin theorem) and
$\bigcup_{\alpha<\omega_1}{\bf\Sigma}^{-1,\theta}_\alpha(X)=
{\bf\Delta}^0_{\theta+1}(X)$ for all  $0<\theta<\omega_1$
(Hausdorff-Kuratowski theorem). In \cite{br} these theorems were
also extended to the quasi-Polish spaces.

\section{Acceptability and Precompleteness}\label{precomp}

Principal TRs from Section \ref{princip} have a property
similar  to the corresponding property  (of being principal
computable) of the standard numbering of the computably enumerable
sets. In this section we establish some other such properties of
the principal $\Gamma$-TRs, namely those of acceptability and precompleteness.

For any set $S$, call a TR
$\nu:\mathcal{N}\to S^\mathcal{N}$ {\em acceptable} if $rng(\nu)$
is downward closed under $\leq$, $\bigsqcup\limits_a\nu_a\in
rng(\nu)$ (where $(\bigsqcup\limits_a\nu_a)\langle
a,b\rangle=\nu_a(b)$) and $\nu_a\langle b,c\rangle =\nu_{s\langle
a,b\rangle} (c)$ for some continuous function $s$ on
$\mathcal{N}$. Here $\nu_a$ is identified with $\nu(a)$ This
definition applies to TRs of the form $\nu:\mathcal{N}\to
P(\mathcal{N})$ if we identify $2^\mathcal{N}$ with
$P(\mathcal{N})$ as in the beginning of Section \ref{naming}.

 \begin{prop}\label{p-in-nu1}\hfill
 \begin{enumerate}[\em(1)]
 \item Any two acceptable TRs of the same subset of
$S^\mathcal{N}$ are equivalent.
 \item If $\mu\equiv\nu$ and $\nu$ is acceptable
then so is $\mu$.
 \end{enumerate}
\end{prop}

\proof 1.  Let $\mu,\nu:\mathcal{N}\to S^\mathcal{N}$ be
acceptable TRs with the same range. We  show $\mu\leq\nu$, the
reduction $\nu\leq\mu$ holds then by symmetry. The TR
$\bigsqcup\mathstrut_b\,\mu_b$ is in $rng(\mu)$, hence
$\bigsqcup\mathstrut_b\,\mu_b=\nu_a$ for some $a$. Then $\mu_b(c)=\nu_a\langle
b,c\rangle  =\nu_{s\langle a,b\rangle }(c)$, hence the continuous
function $b\mapsto s\langle a,b\rangle$ reduces $\mu$ to $\nu$.

2. Straightforward.
\qed

\noindent Next  we show that the principal TRs of pointclasses in
$\mathcal{N}$ are acceptable.

 \begin{prop}\label{p-in-nu2}
Let $\Gamma$ be a family of pointclasses  such that
$\Gamma(\mathcal{N})$ has a principal $\Gamma$-TR $\nu$. Then
$\nu$ is acceptable.
 \end{prop}

\proof By the definition of a family of pointclassed,
$rng(\nu)$ is downward closed under $\leq$. The property
$\bigsqcup\limits_a\nu_a\in rng(\nu)$ holds because
$U_\nu\in\Gamma(\mathcal{N}\times\mathcal{N})$,
 $\langle a,b\rangle\in\bigsqcup\mathstrut_a\,\nu_a\leftrightarrow (a,b)\in U_\nu$, and
 $\mathcal{N}\times\mathcal{N}$ is homeomorphic to $\mathcal{N}$.

The TR $\mu$ of $\Gamma(\mathcal{N})$  defined by $\mu\langle
a,b\rangle=\{c\mid (a,\langle b,c\rangle)\in U_\nu\}$, is a
$\Gamma$-TR, hence $\mu\leq\nu$ via a continuous function $s$ on
$\mathcal{N}$. Then
 $$c\in\nu_{s\langle a,b\rangle}\leftrightarrow
  c\in\mu{\langle a,b\rangle}\leftrightarrow
  \langle b,c\rangle\in\nu_a.$$
 In the
``characteristic functions'' notation this means exactly
$\nu_a\langle b,c\rangle =\nu_{s\langle a,b\rangle} (c)$, hence
$\nu$ is acceptable.
\qed

From Theorem \ref{prin} we now immediately obtain:

 \begin{cor}\label{p-in-nu3}
Let $\Gamma$ be an
arbitrary non-self-dual level of a hierarchy from Section
\ref{hier}. Then the principal TR of $\Gamma(\mathcal{N})$ is acceptable.
 \end{cor}

Next we show that the principal TRs  of the non-self-dual levels
of the classical hierarchies are precomplete. The notion of
precompeteness is very important in the numbering theory
\cite{er77}. In \cite{wei87} the theory of precomplete numberings
was extended to the context of representations where, as usual,
the theory splits to the ``computable'' and ``topological''
versions. Here we consider only the topological version.

Recall from Chapter  3 of \cite{wei87} that a TR $\nu$ is {\em
precomplete} if for any partial continuous function $\psi$ on
$\mathcal{N}$ there is a total continuous function $g$ on
$\mathcal{N}$ that extends $\psi$ modulo $\nu$, i.e.
$\nu\psi(x)=\nu g(x)$ whenever $\psi(x)$ is defined (we call $g$ a
{\em $\nu$-totalizer} of $\psi$). Precomplete TRs have several nice
properties, in particular they satisfy the recursion theorem and
the Rice theorem. The recursion theorem for a TR $\nu$ means the
uniform  fixed point property (FPP).

We say that a TR $\nu$ {\em has FPP} if for any continuous
function $f$ on $\mathcal{N}$ there is $c\in\mathcal{N}$ (a {\em
fixed point} of $f$ w.r.t. $\nu$) such that $\nu(c)=\nu f(c)$. The
uniform FPP means that the fixed point $c$ may be found
continuously from a given index for $f$ in a natural TR $\phi$ of
all partial continuous function $\psi$ on $\mathcal{N}$ with a
$\mathbf{\Pi}^0_2$-domain (more formally, there is a  continuous
function $c$ on $\mathcal{N}$ such that $\nu(c(x))=\nu
\phi_x(c(x))$ whenever $\phi_x$ is total).

We show that the precompleteness property is preserved by the
operations on families  of pointclasses in Section \ref{spaces}.
In the next lemma we use the notation of Lemma \ref{presprin}.

 \begin{lem}\label{precomp1}\hfill
 \begin{enumerate}[\em(1)]
 \item Let $\mathcal{A}\subseteq P(\omega)$ and let $\Gamma$ be a family of
pointclasses such that  $\Gamma(X)$ has a principal $\Gamma$-TR
$\nu$ which is precomplete. Then the TR $\nu_\mathcal{A}$ of
$\Gamma_\mathcal{A}(X)$ is precomplete.
 \item If $\Gamma$ is a family of
pointclasses  such that $\Gamma(\mathcal{N}\times X)$ has a
principal $\Gamma$-TR  $\nu$ which is  precomplete then the TR
$\nu_p$ of $\Gamma_p(X)$ is precomplete.
 \end{enumerate}
\end{lem}

\proof 1. Let $\psi$ be a partial continuous function on
$\mathcal{N}$. For any $k<\omega$, let $p_k$ be the continuous
function on $\mathcal{N}$ such that $p_k(\langle
a_0,a_1,\ldots\rangle)=a_n$ for all $a_i\in\mathcal{N}$. For any
$n<\omega$, let $g_n$ be a $\nu$-totalizer of $p_n\circ\psi$. Then
the continuous function $g(x)=\langle g_0(x),g_1(x),\ldots\rangle$
is a  $\nu_\mathcal{A}$-totalizer of $\psi$. Therefore,
$\nu_\mathcal{A}$ is precomplete.

2. Let $\psi$ be a partial continuous function on $\mathcal{N}$,
then there is a continuous $\nu$-totalizer $g$ of $\psi$. By the
definition of $\nu_p$, $g$ is also a $\nu_p$-totalizer of $\psi$.
Therefore, $\nu_p$ is precomplete.
\qed

The following main result of this section shows that all principal TRs of
non-self-dual levels of the  hierarchies from Section \ref{hier}
are precomplete.

\begin{thm}\label{precomp2}
Let $X$ be a countably based space and let $\Gamma$ be an
arbitrary non-self-dual level of a hierarchy from Section
\ref{hier}. Then the principal $\Gamma$-TR of $\Gamma(X)$ is precomplete.
\end{thm}

\proof We consider first the principal TR $\pi$ of
$\mathbf{\Sigma}^0_1(X)$  defined in the proof of Theorem
\ref{prin}.  We have to show that $\pi$ is precomplete. Let $\psi$
be a partial continuous function on $\mathcal{N}$, we have to find
a continuous $\pi$-totalizer $g$ of $\psi$. As is well known
\cite{wei87,wei92}, we can without loss of generality think that
$\psi=\phi_a$ for some $a\in\mathcal{N}$, i.e., for each
$x\in\mathcal{N}$, $\psi(x)=\varphi_n^{a\oplus x}$ is the $n$-th
(where $n=a(0)$) partial computable function on $\omega$ with
oracle $a\oplus x$. (Here we use standard notation from
computability theory.) It is straightforward to define a
continuous function $g$ on $\mathcal{N}$ with the following
properties:

\begin{iteMize}{$\bullet$}
 \item if $0\not\in dom\psi(x)$, then $g(x)=0^\omega$;
 \item if $0\in dom\psi(x)$ but $1\not\in dom\psi(x)$, then
$g(x)=0^{i_0}n_00^\omega$ for some $i_0<\omega$, where
$n_0=\psi(x)(0)$;
 \item if $0,1\in dom\psi(x)$ but $2\not\in dom\psi(x)$, then
$g(x)=0^{i_0}n_00^{i_1}n_10^\omega$ for some $i_0,i_1<\omega$,
where $n_0=\psi(x)(0)$ and $n_1=\psi(x)(1)$;

$ .\;\;.\;\;.\;\;.\;\;.\;\;.\;\;.\;\;.$

 \item if $dom\psi(x)=\omega$  then
$g(x)=0^{i_0}n_00^{i_1}n_10^{i_2}n_2\cdots$ for some
$i_0,i_1,\ldots<\omega$, where $n_i=\psi(x)(i)$ for each
$i<\omega$.
 \end{iteMize}

\noindent From the definition of $\pi$ it follows that $g$ is a
$\pi$-totalizer of $\psi$.

For the other levels the assertion follows from Lemmas
\ref{hierop}  and \ref{precomp1} because, as is well known
\cite{wad84},  any non-self-dual level of the Borel hierarchy and
of the difference hierarchies coincides with $O_\mathcal{A}(X)$
for some Borel set $\mathcal{A}\subseteq P(\omega)$.
\qed

As is well known, precompleteness  implies the Rice theorem. In
particular, for the principal TR $\pi$ of the open sets the Rice
theorem looks as follows:

 \begin{prop}\label{rice}
 Let $X$ be a countably based space and
$\mathcal{A}\subseteq\mathbf{\Sigma}^0_1(X)$. Then
$\pi^{-1}(\mathcal{A})\in\mathbf{\Delta}^0_1(\mathcal{N})$ iff
$\mathcal{A}=\emptyset$ or $\mathcal{A}=\mathbf{\Sigma}^0_1(X)$.
 \end{prop}

\proof We consider  the implication from left to right
because  implication in the opposite direction is obvious. Let
$\pi^{-1}(\mathcal{A})\in\mathbf{\Delta}^0_1(\mathcal{N})$ and
suppose for a contradiction that
$\emptyset\subset\mathcal{A}\subset\mathbf{\Sigma}^0_1(X)$, so
$a\in\pi^{-1}(\mathcal{A})\not\ni b$ for some $a,b\in\mathcal{N}$.
Let $f$ be the function on $\mathcal{N}$ that sends
$\pi^{-1}(\mathcal{A})$ to $b$ and the complement of
$\pi^{-1}(\mathcal{A})$ to $a$. Since $\pi^{-1}(\mathcal{A})$ is
clopen, $f$ is continuous. By FPP for the precomplete TR $\pi$,
$\pi(c)=\pi f(c)$ for some $c$, but this is contradictory.
\qed

For some other  levels of the classical hierarchies the Rice
theorem have interesting modifications, in particular we have:

 \begin{prop}\label{rice2}
 Let $X$ be a countably based space,
$\mathcal{A}\subseteq\mathbf{\Sigma}^{-1}_2(X)$,  and let $\nu$ be
a principal $\mathbf{\Sigma}^{-1}_2$-TR of
$\mathbf{\Sigma}^{-1}_2(X)$. Then
$\nu^{-1}(\mathcal{A})\in\mathbf{\Delta}^{-1}_2(\mathcal{N})$ iff
$\mathcal{A}=\emptyset$ or $\mathcal{A}=\mathbf{\Sigma}^{-1}_2(X)$.
 \end{prop}

\proof We consider  the implication from left to right
because  implication in the opposite direction is obvious. Let
$\nu^{-1}(\mathcal{A})\in\mathbf{\Delta}^{-1}_2(\mathcal{N})$  and
suppose for a contradiction that
$\mathcal{A}\not=\{\emptyset,\mathbf{\Sigma}^{-1}_2(X)\}$. We may
assume without loss of generality that
$\emptyset\not\in\mathcal{A}$ (otherwise, replace $\mathcal{A}$ by
$\mathbf{\Sigma}^{-1}_2(X)\setminus\mathcal{A}$). Let
$C\in\mathcal{A}$.

Let $A_0,A_1\in\mathbf{\Sigma}^0_1(\mathcal{N})$  satisfy
$\nu^{-1}(\mathcal{A})=A_0\setminus A_1$ and  $A_0\supseteq A_1$.
Let $C_0,C_1\in\mathbf{\Sigma}^0_1(X)$ satisfy $C=C_0\setminus
C_1$ and  $C_0\supseteq C_1$. From the definition of $\pi$ it is
straightforward to find continuous functions $f_0,f_1$ on
$\mathcal{N}$ such that: $\pi f_0(x)=\emptyset$ for
$x\in\mathcal{N}\setminus A_0$ and $\pi f_0(x)=C_0$ for $x\in
A_0$; $\pi f_1(x)=C_1$ for $x\in\mathcal{N}\setminus A_1$ and $\pi
f_1(x)=C_0$ for $x\in A_1$. Finally, let $f(x)=\langle
f_0(x),f_1(x)\rangle$, then $\nu f(x)=\pi f_0(x)\setminus \pi
f_1(x)$.

Note that  $x\not\in A_0$ implies $\nu f(x)=\emptyset$,   $x\in
A_0\setminus A_1$ implies $\nu f(x)=C$, and  $x\in A_1$ implies $\nu
f(x)=\emptyset$. Altogether, $x\in\nu^{-1}(\mathcal{A})$ iff
$f(x)\not\in\nu^{-1}(\mathcal{A})$. By Theorem \ref{precomp2},
$\nu$ has the FPP-property, i.e. $\nu(c)=\nu f(c)$ for some
$c\in\mathcal{N}$. Then $c\in\nu^{-1}(\mathcal{A})$ iff
$c\not\in\nu^{-1}(\mathcal{A})$, a contradiction.
\qed

\section{Principal Continuous Total Representations}\label{princon}

Working with a space $X$, it is natural to look at continuous TRs
of $X$, hence it is instructive to ask for which spaces a
principal TR  in the class of continuous TRs exists. We call a TR
$\gamma$ of a space $X$ {\em principal} if it is continuous, and
any continuous TR $\nu:\mathcal{N}\to X$ is reducible to $\gamma$.
In this section we show that principal continuous TRs share some
basic properties of admissible representations
\cite{wei00,sch02,sch03}. Our proofs are easy adaptations of the
well known corresponding proofs for admissible representations.

We start with recalling some  properties of sequential topologies.
Let $X$ be an arbitrary set.  By a topology on $X$ we mean the
corresponding class of open sets.  Let $\mathcal{T}(X)$ be the set
of all topologies on $X$, and let $\tau\in\mathcal{T}(X)$.  A
sequence $\{x_n\}$ in $X$  {\em $\tau$-converges} to an element
$x\in X$ if for any $U\in\tau$, the condition $x\in U$ implies
that $x_n$ is eventually in $U$ (i.e., there is $n_0<\omega$ such
that $x_n\in U$ for all $n\geq n_0$). Let $\tau^s$ be the set of
all $A\subseteq X$ such that for all $x,x_0,x_1,\ldots\in X$, if
$\{x_n\}$ $\tau$-converges to $x$ and $x\in A$ then $x_n$ is
eventually in $A$. Note that our notation $\tau^s$ corresponds to
notation $seq(\tau)$ in \cite{sch02}. The next two lemmas follow
from  well known facts in \cite{en89,sch02}.

\begin{lem}\label{seq}
For any set $X$,  $\tau\mapsto\tau^s$ is a closure operation on
$(\mathcal{T}(X);\subseteq)$,  i.e. $\tau^s\in\mathcal{T}(X)$ for
$\tau\in\mathcal{T}(X)$, $\tau\subseteq\tau^s$,
$(\tau^s)^s=\tau^s$, and $\tau\subseteq\tau_1$ implies
$\tau^s\subseteq\tau^s_1$.
\end{lem}

For a space $X$, let $\tau_X$  denote the  topology  on $X$ (i.e.
$\tau_X=\mathbf{\Sigma}^0_1(X)$). A function $f:X\to Y$ between
spaces is {\em sequentially continuous} if for all
$x,x_0,x_1,\ldots\in X$ such that $\{x_n\}$    $\tau_X$-converges
to  $x$, $\{f(x_n)\}$    $\tau_Y$-converges to $f(x)$.

\begin{lem}\label{seq3}\hfill
\begin{enumerate}[\em(1)]
 \item Any continuous function is sequentially continuous.
 \item If $X$ is countably based then $\tau_X=\tau^s_X$.
 \item If $f:X\to Y$ is sequentially continuous and $X$
is sequential (in particular, countably based) then $f$ is continuous.
 \end{enumerate}
\end{lem}

\noindent The next result  is a slight modification of the
corresponding assertion for the admissible representations
\cite{sch02}. For a TR $\gamma$ of a set $X$, let
$\tau_\gamma$ denote the {\em final topology of $\gamma$} on $X$
consisting of all sets $A\subseteq X$ such that
$\gamma^{-1}(A)\in\mathbf{\Sigma}^0_1(\mathcal{N})$

\begin{thm}\label{seq1}
Let $\gamma$ be a principal continuous TR of a space $X$.
 \begin{enumerate}[\em(1)]
 \item Any principal continuous TR of $X$ is equivalent to $\gamma$.
 \item $X$ is a $T_0$-space.
 \item For all $x,x_0,x_1,\ldots\in X$, $\{x_n\}$  $\tau_X$-converges
to  $x$ iff there exist
$a\in\gamma^{-1}(x),a_0\in\gamma^{-1}(x_0),a_1\in\gamma^{-1}(x_1),\ldots$
such that $\{a_n\}$  $\tau_\mathcal{N}$-converges to  $a$.
 \item $\tau_\gamma$ is the sequentialization of $\tau_X$,
 i.e. $\tau^s_X=\tau_\gamma$.
 \end{enumerate}
\end{thm}

\proof 1. Obvious.

2. Suppose not,  so there are distinct $x,y\in X$ such that
$\forall U\in\tau_X(x\in U\leftrightarrow y\in U)$. Then any
function $\nu:\mathcal{N}\to\{x,y\}$ is continuous. Since $\gamma$
is principal, $\nu\leq\gamma$ for all such $\nu$. But this is not
possible because there are hypercontinuum many of such
$\nu:\mathcal{N}\to\{x,y\}$ and only continuum many of $\nu$
reducible to $\gamma$ (because there are only continuum many
continuous functions on $\mathcal{N}$).

3. Let $\{x_n\}$ $\tau_X$-converges to  $x$. Clearly,
$A=\{0^\omega,0^n1^\omega\mid n<\omega\}$ is a retract of
$\mathcal{N}$, i.e. for some continuous function $r:\mathcal{N}\to
A$ we have $r(a)=a$ for all $a\in A$. Moreover, $\{0^n1^\omega\}$
$\tau_\mathcal{N}$-converges to  $0^\omega$, hence the function
$g:A\to X$ is continuous where $g(0^\omega)=x$ and
$g(0^n1^\omega)=x_n$ for all $n<\omega$. Since $g\circ
r:\mathcal{N}\to X$ is continuous and $\gamma$ is principal,
$g\circ r=\gamma\circ f$ for some continuous function $f$ on
$\mathcal{N}$. Then elements $a=f(0^\omega)$, $a_n=f(0^n1^\omega)$
have the desired properties because $\{a_n\}$
$\tau_\mathcal{N}$-converges to  $a$ by continuity of $f$,
$\gamma(a)=\gamma f(0^\omega)=gr(0^\omega)=g(0^\omega)=x$, and
similarly $\gamma(a_n)=x_n$ for each $n<\omega$.

Conversely,  let $a,a_0,a_1,\dots$ have the specified properties,
we have to check that $\{x_n\}$  $\tau_X$-converges to  $x$. Let
$x\in U\in\tau_X$, then $a\in\gamma^{-1}(U)$, and
$\gamma^{-1}(U)\in\tau_\mathcal{N}$ by continuity of $\gamma$.
Since $\{a_n\}$  $\tau_\mathcal{N}$-converges to  $a$, $a_n$ is
eventually in $\gamma^{-1}(U)$. Therefore, $x_n$ is eventually in
$U$.

4. Let $A\in\tau^s_X$.  We have to show that $A\in\tau_\gamma$,
i.e. $\gamma^{-1}(A)\in\tau_\mathcal{N}$, i.e. for any
$a\in\gamma^{-1}(A)$ there is $n<\omega$ with
$a[n]\cdot\mathcal{N}\subseteq\gamma^{-1}(A)$. Suppose for
contradiction that there is $a\in\gamma^{-1}(A)$ such that
$a[n]\cdot\mathcal{N}\not\subseteq\gamma^{-1}(A)$  for all
$n<\omega$. For any $n<\omega$, choose $a_n\in
a[n]\cdot\mathcal{N}\setminus\gamma^{-1}(A)$.  Then $\{a_n\}$
$\tau_\mathcal{N}$-converges to  $a$. By continuity of $\gamma$,
$\{\gamma(a_n)\}$  $\tau_X$-converges to  $\gamma(a)$. Since
$A\in\tau^s_X$ and $\gamma(a)\in A$, $\gamma(a_n)$ is
eventually in $A$, hence  $\{a_n\}$ is eventually in
$\gamma^{-1}(A)$. A contradiction.

Conversely, let $A\in\tau_\gamma$, i.e.
$\gamma^{-1}(A)\in\tau_\mathcal{N}$. Let $\{x_n\}$
$\tau_X$-converge to $x\in A$; we have to show that   $x_n$ is
eventually in $A$. Choose $a,a_0,a_1,\ldots$ as in  item 3, so in
particular $\{a_n\}$  $\tau_\mathcal{N}$-converges to
$a\in\gamma^{-1}(A)$. Then $a_n$ is eventually in
$\gamma^{-1}(A)$, so $x_n=\gamma(a_n)$ is eventually in $A$.
\qed

The next  important property of principal continuous TRs is
again analogous to the corresponding property of admissible
representations.

\begin{thm}\label{seq2}
Let $\gamma$ and $\delta$ be  principal  continuous TRs of
spaces $X$ and $Y$, respectively. Then $f:X\to Y$ is sequentially
continuous iff there exists a continuous function
$\hat{f}:\mathcal{N\to}\mathcal{N}$ with
$f\circ\gamma=\delta\circ\hat{f}$. In particular,
$A\mapsto\gamma^{-1}(A)$ is a homomorphism from $(P(X);\leq_W)$
into $(P(\mathcal{N});\leq_W)$.
\end{thm}

\proof Let $f$ be sequentially  continuous, then so is also
$f\circ\gamma$. By Lemma \ref{seq3}, $f\circ\gamma$ is continuous.
Since $\delta$ is principal, $f\circ\gamma=\delta\circ\hat{f}$ for
some continuous function $\hat{f}$ on $\mathcal{N}$.

Conversely, let $\hat{f}$ be continuous  with the specified
property, and let $\{x_n\}$    $\tau_X$-converge  to $x$. Choose
$a,a_0,a_1,\ldots$ as in item 3 of Proposition \ref{seq1}, so in
particular $\{a_n\}$    $\tau_\mathcal{N}$-converges to  $a$.
Since $\delta\circ\hat{f}$ is continuous, $\{\delta\hat{f}(a_n)\}$
$\tau_Y$-converges to  $\delta\hat{f}(a)$, hence also
$\{f\gamma(a_n)\}$    $\tau_Y$-converges to  $f\gamma(a)$.
Therefore $\{f(x_n)\}$    $\tau_Y$-converges to  $f(x)$, as
desired.
\qed

\begin{rems}\hfill
\begin{enumerate}[1.]
\item.  In numbering theory, a partial analogy to
principal continuous TRs is provided by the so called approximable
numberings \cite{er77,s06}.

\item We see that some  important properties of principal continuous
TRs are close to those of admissible representations which are
very popular in CA. Obviously, every principal continuous TR of a
space X that admits an admissible TR is already admissible.  Also,
every admissible TR is principal continuous. Unfortunately,
currently we do not know whether the converse implication is also
true. If yes, this would be a new interesting characterization of
the admissible TRs (and we believe the results of this section
could be useful to prove this). If no, we would obtain a  new
concept of interest for CA. In the next section  we continue to
discuss the relationships between admissible and principal
continuous TRs.

\end{enumerate}
\end{rems}

\section{Admissible Total Representations}\label{admiss}

A fundamental notion of CA is the notion of admissible
representation, i.e. (in terminology of Section \ref{princip}),
principal continuous representations. This notion was introduced
in \cite{kw85} for countably based spaces and it was extensively
studied by many authors. In \cite{bh02} a close relation of
admissible representations of countably based spaces to open
continuous representations was established. In \cite{sch02} the
notion was extended to non-countably based spaces and a nice
characterization of the admissibly represented spaces was
achieved. In \cite{sch95,sch04} the admissible representations
allowing a computational complexity theory in CA were identified.

As mentioned above, the previous study of admissible
representations in CA paid no attention to TRs which was in a
striking contrast with numbering theory where total numberings
obviously dominate. But recently it became clear that the
admissible TRs deserve more attention. Recall that a
representation  $\alpha$ of a space $X$ (i.e., a partial
surjection from  $\mathcal{N}$ onto $X$) is {\em admissible} if it
is continuous and any partial continuous function $\phi$ from
$\mathcal{N}$ to $X$ is reducible to $\alpha$ (i.e., there is a
partial continuous function $f$ on $\mathcal{N}$ such that
$\phi(x)=\alpha f(x)$ for each $x\in dom(f)$).

 \begin{prop}\label{pr-adm}
 If a space has a principal continuous TR then it has an
 admissible partial representation.
 \end{prop}

\proof Let $\gamma$ be a principal continuous  TR of a space
$X$. By Theorem 13 in \cite{sch02}, it suffices to show that $X$
is a $T_0$-space which has a countable pseudobase. The $T_0$
property holds by item 2 of Theorem \ref{seq1}.  A countable
pseudobase for $X$ may be constructed similarly to Lemma 11 in
\cite{sch02}. Namely, let $\mathcal{B}$ be a countable base for
$\mathcal{N}$ (say,
$\mathcal{B}=\{\sigma\cdot\mathcal{N}\mid\sigma\in\omega^\ast\}$);
we check that $\{\gamma(B)\mid B\in\mathcal{B}\}$ is a countable
pseudobase for $X$. This by definition means that if $\{x_n\}$
$\tau_X$-converges to $x\in U$ for some $U\in\tau_X$ then there is
$B\in\mathcal{B}$ such that $\gamma(B)\subseteq U$,
$x\in\gamma(B)$, and $x_n$ is eventually in $\gamma(B)$ (i.e.,
there is $n_0<\omega$ such that $x_n\in\gamma(B)$ for each $n\geq
n_0$).

By item 3 of Theorem \ref{seq1}, there exist
$a\in\gamma^{-1}(x),a_0\in\gamma^{-1}(x_0),a_1\in\gamma^{-1}(x_1),\ldots$
such that $\{a_n\}$  $\tau_\mathcal{N}$-converges to $a$. Since
$\gamma(a)=x\in U$ and $\gamma$ is continuous,
$a\in\gamma^{-1}(U)\in\tau_\mathcal{N}$. Since $\mathcal{B}$ is a
base for $\mathcal{N}$, $a\in B\subseteq\gamma^{-1}(U)$  for some
$B\in\mathcal{B}$. Then $a_n$ is eventually in $B$, hence $x_n$ is
eventually in $\gamma(B)\subseteq U$, and
$x=\gamma(a)\in\gamma(B)$.
\qed

From the recent paper of M. de Brecht \cite{br}
it follows that admissible TRs are sufficient for treating a
large and useful class of countably based spaces. The following
assertion is contained among results in  \cite{br}, we only slightly
reformulate it in order to put emphasis on total rather than
partial representations.

 \begin{prop}\label{adm} For any countably based space $X$
the following statements are equivalent:
 \begin{enumerate}[\em(1)]
 \item $X$ is quasi-Polish.
 \item $X$ has an open continuous TR.
 \item $X$ has an admissible TR.
 \end{enumerate}
\end{prop}

\noindent{\bf Proof Sketch.} 1$\to$2. We reproduce  the short proof
from \cite{br}. By Proposition \ref{pi2} we may assume that $X$ is
a $\mathbf{\Pi}^0_2$-subset of $P\omega$. The equation
$\rho(a)=\{n\in\omega\mid n+1\in rng(a)\}$ defines an open
continuous TR of $P\omega$. Its restriction $\rho^\prime$ to
$\rho^{-1}(X)$ is an open continuous surjection from the subspace
$\rho^{-1}(X)$ of $\mathcal{N}$ onto $X$. Since $\rho^{-1}(X)$ is
in $\mathbf{\Pi}^0_2(\mathcal{N})$, it is a Polish space by
Theorem 3.11 in \cite{ke94}. By Exercise 7.14 in \cite{ke94}, there is
an open continuous TR $f$ of $\rho^{-1}(X)$. Then
$\rho^\prime\circ f$ is an open continuous TR of $X$.

2$\to$3. From (the proof of) Theorem 12 in \cite{bh02} it follows
that any open continuous TR is admissible.

3$\to$1. A non-trivial result in \cite{br}.
\qed

\begin{rems}
\begin{enumerate}[1.]
\item
 Note that  any open continuous TR of $X$ is
automatically admissible but the converse does not hold in general
\cite{bh02}.

\item From \cite{sch02}  we know that sequential   admissibly
represented spaces form a cartesian closed category but, since
they contain all countably based spaces, many of them have poor
DST-properties  (e.g., they in general do not satisfy the
Hausdorff-Kuratowski theorem). From \cite{br} we know that the
countably based admissibly totally  representable spaces (i.e.,
the quasi-Polish spaces) have good DST-properties but, as recently
M. Schr\"oder has shown answering to my question, they do not form
a cartesian closed category. It seems that combining  the both
properties (of being cartesian closed and having a good DST) is
not possible for large enough classes of spaces. To my knowledge,
only some rather small classes of domains are known to have both
properties.

\item Let us stress that the  question whether there is any space
that admits a principal continuous TR but not an admissible TR
remains open.
\end{enumerate}
\end{rems}

\noindent An advantage of admissible TRs (compared with partial admissible
representations) is that index sets of TRs behave more
``regularly'' than those of the partial representations; we
discuss this in more detail in  Section  \ref{categ}.  Another
advantage  is that it is a ``more canonical'' notion. We
illustrate this by providing easy topological invariants for the
quasi-Polish spaces in terms of admissible TRs.

 \begin{prop}\label{inv}
Let $\alpha,\beta$ be admissible TRs of quasi-Polish spaces
$X,Y$, respectively. Then the kernel relation (see Section \ref{naming})
$E_\alpha$ is in $\mathbf{\Pi}^0_2(\mathcal{N})$, and $E_\alpha\equiv_W
E_\beta$ whenever $X$ and $Y$ are homeomorphic. In particular, the
Wadge degree of $E_\alpha$ is a topological invariant of $X$.
\end{prop}

\proof The equality relation on $X$ (as well as on arbitrary
countably based $T_0$-space \cite{br}) is in
$\mathbf{\Pi}^0_2(X\times X)$. Since $\langle a,b\rangle\in
E_\alpha$ iff $\alpha(a)=\alpha(b)$, $E_\alpha$ is a continuous
preimage of the equality relation, hence
$E_\alpha\in\mathbf{\Pi}^0_2(\mathcal{N})$.

For the second assertion, assume that $X,Y$ are homeomorphic. Then
$\alpha\equiv\beta$, hence $E_\alpha\equiv_W E_\beta$.
\qed

As is well known, the structure of Wadge degrees of
$\mathbf{\Pi}^0_2(\mathcal{N})$ sets is very simple, namely it
precisely corresponds to the Wadge complete sets in levels
$\mathbf{\Sigma}^{-1}_\alpha$,$\mathbf{\Pi}^{-1}_\alpha$,
$\mathbf{\Delta}^{-1}_{1+\alpha}$ ($\alpha<\omega_1$) of the
difference hierarchy over $\mathbf{\Sigma}^0_1(\mathcal{N})$, plus
the Wadge degree of a $\mathbf{\Pi}^0_2$-complete set. Hence, the
previous Proposition suggests a natural classification of
quasi-Polish spaces $X$ according to the mentioned level in which
$E_\alpha$ is Wadge complete.
Obviously, $E_\alpha$ cannot be in
$\mathbf{\Sigma}^{-1}_0=\{\emptyset\}$ (provided that we do not
consider the empty space), $E_\alpha\in\mathbf{\Pi}^{-1}_0$ iff
$X$ is a singleton space, $E_\alpha$ is Wadge complete in
$\mathbf{\Delta}^{-1}_1=\mathbf{\Delta}^0_1$ iff $X$ is a
non-singleton discrete space (hence $X$ is at most countable), and
$E_\alpha\in\mathbf{\Sigma}^{-1}_1=\mathbf{\Sigma}^0_1$ implies
$E_\alpha\in\mathbf{\Delta}^0_1$. It is a nice open question to
precisely characterize the quasi-Polish spaces in any class of
this classification.

Although there is no such elegant classification of arbitrary
admissibly represented spaces, one can use a slightly more
complicated invariant. For a (partial) representation
$\delta$ of a space $X$, let
$E_\delta=\{\langle a,b\rangle\mid a,b\in
dom(\delta)\wedge\delta(a)=\delta(b)\}$. Let $w(X)$ be the set of
minimal Wadge degrees that contain $E_\delta$ for some admissible
representation $\delta$ of $X$. From the structure of Wadge
degrees it follows that (at least, under some more or less
reasonable set-theoretic axioms) $w(X)$ always exists and it
consists either of one or of two elements. A natural question is
to characterize the range of the function $w$.

{\bf Remark.} Note  that the aforementioned classification of
topological spaces is related to the separability axioms, in
particular $E_{\alpha_X}\in\mathbf{\Pi}^0_1(\mathcal{N})$ for any
Hausdorff space $X$ (where $\alpha_X$ is an admissible TR of $X$).
One could also measure the complexity of $X$ by the complexity of
singletons $\{x\}$ for $x\in X$ (or by the complexity of their
index sets $\alpha_X^{-1}(\{x\})$). E.g., $X$ is a $T_1$-space iff
$\{x\}\in\mathbf{\Pi}^0_1(X)$ for each $x\in X$,  $X$ is a
$T_D$-space \cite{br} iff $\{x\}\in\mathbf{\Sigma}^{-1}_2(X)$ for
each $x\in X$, and if $X$ is a countably based $T_0$-space then
any singleton set is in $\mathbf{\Pi}^0_2(X)$ \cite{br}.  Note
that the last complexity measure (suggested by a referee) is
related to the first one because any singleton set is a continuous
preimage of the equality relation on $X$.

Another instructive question related to Proposition \ref{adm}
is to investigate  ``natural'' non-countably based spaces having an
admissible TR. To see that such spaces exist consider again
the principal TR $\pi$ of $\mathbf{\Sigma}^0_1(X)$ for a
countably based space $X$ (see the proof of Theorem \ref{prin}).

There are at least two natural topologies on
$\mathbf{\Sigma}^0_1(X)$, for an arbitrary space $X$. First, this is
the Scott topology $\sigma$ on the complete lattice
$(\mathbf{\Sigma}^0_1(X);\subseteq)$. Recall that
$\mathcal{A}\in\sigma$ iff $\mathcal{A}$ is upward closed in
$(\mathbf{\Sigma}^0_1(X);\subseteq)$, and $\mathcal{D}\cap
\mathcal{A}\not=\emptyset$ for each   directed subset  $\mathcal{D}$
of $(\mathbf{\Sigma}^0_1(X);\subseteq)$ with
$\bigcup\mathcal{D}\in\mathcal{A}$. Second, this is the compact-open
topology $\kappa$, the basic open sets of which are of the form
$O(K)=\{A\in\mathbf{\Sigma}^0_1(X)\mid K\subseteq A\}$ where $K$
runs through the compact subsets of $X$. For Polish spaces $X$ the
topologies $\sigma$ and $\kappa$ are known to coincide. In the
general case we have:

 \begin{prop}\label{scott-comp}\hfill
Let  $X$ be an arbitrary topological space.
 \begin{enumerate}[\em(1)]
 \item $\kappa\subseteq\sigma$.
 \item If $\{A_n\}$ $\kappa$-converges to $A\in\mathbf{\Sigma}^0_1(X)$ and
$K\subseteq A$ for some compact subset $K$  of $X$ then eventually
$K\subseteq A_n$.
 \end{enumerate}
 \end{prop}

\proof 1. It suffices to show that $O(K)\in\sigma$ for each
compact  $K\subseteq X$. Clearly, $O(K)$ is upward closed in
$(\mathbf{\Sigma}^0_1(X);\subseteq)$. It remains to show that if
$\bigcup\mathcal{D}\in O(K)$ where $\mathcal{D}$ is a directed
subset of $(\mathbf{\Sigma}^0_1(X);\subseteq)$ then
$\mathcal{D}\cap O(K)\not=\emptyset$. Let
$K\subseteq\bigcup\mathcal{D}$. Since $K$ is compact, $K\subseteq
D_0\cup\cdots\cup D_n$ for some $n<\omega$ and
$D_0,\ldots,D_n\in\mathcal{D}$. Since $\mathcal{D}$ is directed,
$D_0\cup\cdots\cup D_n\subseteq D$ for some $D\in\mathcal{D}$.
Therefore $D\in\mathcal{D}\cap O(K)$.

2. Since $A\in O(K)$ and $O(K)\in\kappa$, $A_n$ is eventually in
$O(K)$, hence eventually $K\subseteq A_n$.
\qed

\noindent Next we show that in many  cases the TR $\pi$ is admissible (cf.
Propositions 4.4.1 and 4.4.3 in \cite{sch03}).

\begin{thm}\label{admpi}
Let  $X$ be a countably based topological  space. Then $\pi$ is an
admissible TR of both $(\mathbf{\Sigma}^0_1(X);\sigma)$ and
$(\mathbf{\Sigma}^0_1(X);\kappa)$.
\end{thm}

\proof First we show that $\pi$ is continuous.  Since
$\kappa\subseteq\sigma$, it suffices to check that $\pi$ is
continuous with respect to $\sigma$, i.e. that
$\pi^{-1}(\mathcal{A})$ is open in $\mathcal{N}$ for each
$\mathcal{A}\in\sigma$. Let $a\in\pi^{-1}(\mathcal{A})$, i.e.
$\pi(a)\in\mathcal{A}$; we have to show that
$a[n]\cdot\mathcal{N}\subseteq\pi^{-1}(\mathcal{A})$ for some
$n<\omega$. Let $A=\pi(a)=\bigcup_nB_{a(n)}$ and
$A_n=B_{a(0)}\cup\cdots\cup B_{a(n-1)}$ for each $n<\omega$. Then
$A_0\subseteq A_1\subseteq\cdots$ and
$\bigcup_nA_n=A\in\mathcal{A}$. Since $\mathcal{A}\in\sigma$, there
is $n<\omega$ with $A_n\in\mathcal{A}$.  For any $b\in
a[n]\cdot\mathcal{N}$ we then have $A_n\subseteq\pi(b)$, hence
$\pi(b)\in\mathcal{A}$ and $b\in\pi^{-1}(\mathcal{A})$. Thus,
$a[n]\cdot\mathcal{N}\subseteq\pi^{-1}(\mathcal{A})$.

It remains to show that, for each $\tau\in\{\sigma,\kappa\}$, any
$\tau$-continuous function $\nu:D\to\mathbf{\Sigma}^0_1(X)$,
$D\subseteq\mathcal{N}$, is reducible to $\pi$, i.e. $\nu(d)=\pi
f(d)$ for some continuous function $f:D\to\mathcal{N}$. Since
$\kappa\subseteq\sigma$,  it suffices to show this for
$\tau=\kappa$.

First we show the auxiliary assertion that for all $d\in D$ and
$x\in\nu(d)$ (i.e., $(d,x)\in U_\nu$) it holds:
 $$\exists n\in\omega\exists A\in\mathbf{\Sigma}^0_1(X)
 (x\in A\wedge ((d[n]\cdot\mathcal{N})\cap D)\times A\subseteq U_\nu).$$
 Let $\{m\mid x\in B_m\}=\{m_0,m_1,\cdots\}$, then the formula above is equivalent to
 $\exists n (((d[n]\cdot\mathcal{N})\cap D)\times A\subseteq U_\nu)$
where $ A=B_{m_0}\cap\cdots\cap B_{m_n}$.
 Suppose that the auxiliary assertion is false, so
 $\forall n ( ((d[n]\cdot\mathcal{N})\cap D)\times A\not\subseteq U_\nu)$.
 Then there are
$b_0,b_1,\ldots\in\mathcal{N}$ and $x_0,x_1,\ldots\in X$ such that
 $$\forall n\in\omega(d[n]\cdot b_n\in D\wedge x_n\in B_{m_0}\cap\cdots\cap
 B_{m_n}\wedge x_n\not\in\nu(d[n]\cdot b_n)).$$
 Since $\{d[n]\cdot b_n\}$ $\tau_\mathcal{N}$-converges to $d$ and $\nu$
is $\kappa$-continuous,  $\{\nu(d[n]\cdot b_n)\}$ $\kappa$-converges
to $\nu(d)$. Since $x\in\nu(d)\in\mathbf{\Sigma}^0_1(X)$ and
$\{x_n\}$ $\tau_X$-converges to $x$,
$K=\{x,x_{n_0},x_{n_0+1},\ldots\}\subseteq\nu(d)$ for some
$n_0<\omega$. Since $K$ is compact in $X$, by item 2 of Proposition
\ref{scott-comp} there is $n_1<\omega$ such that $\forall n\geq
n_1(K\subseteq\nu(d[n]\cdot b_n))$. Then $x_n\in\nu(d[n]\cdot b_n)$
for all $n\geq n_0,n_1$ which is a contradiction.

Now we define $f:D\to\mathcal{N}$  as follows: $f(d)\langle
i,j\rangle=j$,  if $B_j\subseteq\bigcap\nu(d[i]\cdot\mathcal{N})$,
and $f(d)\langle i,j\rangle=0$ otherwise. Clearly, $f$ is
continuous (even Lipschitz), so it remains to check that
$\nu(d)=\pi f(d)$ for each $d\in D$, i.e.
$\nu(d)=\bigcup\{B_j\mid\exists
i(B_j\subseteq\bigcap\nu(d[i]\cdot\mathcal{N}))\}$. The inclusion
from right to left follows from $d\in d[i]\cdot\mathcal{N}$.
Conversely, let $x\in\nu(d)$. By the auxiliary assertion, there
are $i,j\in\omega$ such that $x\in B_j\subseteq\nu(d[i]\cdot b)$
for all $b\in \mathcal{N}$. Thus, $x$ is in the right hand side of
the equality.
\qed

In particular, $\mathbf{\Sigma}^0_1(\mathcal{N})$ (with any of the
topologies $\sigma,\kappa$) has an admissible TR. As is well known,
the space $\mathbf{\Sigma}^0_1(\mathcal{N}$) is not countably based
(see also Theorem \ref{pi-baire}).  Thus,
$\mathbf{\Sigma}^0_1(\mathcal{N)}$ is a natural example of a
non-countably based admissibly totally representable space.

Let $\tau_\pi$ be the final topology induced by $\pi$ on
$\mathbf{\Sigma}^0_1(X)$.  From Theorem 7 in in \cite{sch02} we
now obtain:

 \begin{cor}\label{open}
Let $X$ be a countably based space.  Then the final topology
$\tau_\pi$ on $\mathbf{\Sigma}^0_1(X)$ coincides with the
sequentialization of any  of the topologies $\sigma,\kappa$, i.e.
$\tau_\pi=\sigma^s=\kappa^s$.
\end{cor}

The last result may be interpreted as a topological analog of the
classical Rice-Shapiro  theorems in computability theory. The next
result was first obtained in \cite{hm82} for the case when $X$ is
Polish.

 \begin{cor}\label{rice-sh}
Let $X$ be a countably based space and
$\mathcal{A}\subseteq\mathbf{\Sigma}^0_1(X)$. Then
$\pi^{-1}(\mathcal{A})\in\mathbf\Sigma^0_1(\mathcal{N})$ iff
$\mathcal{A}\in\kappa^s$.
 \end{cor}

Theorem \ref{admpi} implies that $\pi$ is an admissible TR of
$(\mathbf{\Sigma}^0_1(X);\tau_\pi)$. Can this result be extended
to other principal TRs of levels of the classical hierarchies from
Section \ref{princip}? The answer is no. We prove this  here
only for the class of differences of open sets.

 \begin{prop}\label{no-admis}
 Let $X$ be a countably based space and let $\nu$ be a
principal $\mathbf{\Sigma}^{-1}_2$-TR of
$\mathbf{\Sigma}^{-1}_2(X)$. Then $\nu$ is not admissible w.r.t.
$\tau_\nu$.
  \end{prop}

\proof By Theorem 13 in \cite{sch02}, it suffices to show
that any two points in $\mathbf{\Sigma}^{-1}_2(X)$ are not
separable by sets in $\tau_\nu$ (note that
$\mathbf{\Sigma}^{-1}_2(X)$ has at least two elements for each
non-empty $X$). Suppose the contrary: $A\in\mathcal{A}\not\ni B$
for some $A,B\in\mathbf{\Sigma}^{-1}_2(X)$ and some
$\mathcal{A}\subseteq\mathbf{\Sigma}^{-1}_2(X)$ with
$\nu^{-1}(\mathcal{A})\in\mathbf{\Sigma}^0_1(\mathcal{N})$. By
Proposition \ref{rice2},
$\mathcal{A}\in\{\emptyset,\mathbf{\Sigma}^{-1}_2(X)\}$. A
contradiction.
\qed

\begin{rem} Note that if $X=\{x\}$ is a singleton space then
$(\mathbf{\Sigma}^0_1(X);\tau_\pi)$ is homeomorphic to the
Sierpinski space, while the space
$(\mathbf{\Sigma}^{-1}_2(X);\tau_\nu)$ consists of two points
which are not separable by open sets. Note also that
$(\mathbf{\Sigma}^0_1(\omega);\sigma)$ is homeomorphic to the
domain $P\omega$.
\end{rem}

We conclude  this section with  the following result (suggested by
a referee) stating some interesting properties of the admissible
TR of $\mathbf{\Sigma}^0_1(\mathcal{N})$.

 \begin{thm}\label{pi-baire}
Let  $\pi$ be  the admissible TR of
$\mathbf{\Sigma}^0_1(\mathcal{N})$. Then
$E_\pi\in\mathbf{\Pi}^1_1(\mathcal{N})$,
$\pi^{-1}(\{\mathcal{N}\})$ is Wadge complete in
$\mathbf{\Pi}^1_1(\mathcal{N})$, and the space
$\mathbf{\Sigma}^0_1(\mathcal{N})$ is not countably based.
 \end{thm}

\proof  Let $\sigma_0,\sigma_1,\ldots$ be an enumeration
without repetition of the set $\omega^\ast$ such that $\sigma_0$
is the empty string. Let $\{B_0,B_1,\ldots\}$ be the enumeration
of a base in $\mathcal{N}$ where $B_0=\emptyset$ and
$B_{n+1}=\sigma_n\cdot\mathcal{N}$. We have
 $$\pi(a)\subseteq\pi(b)\leftrightarrow\forall x\in\mathcal{N}
 \forall n\in\omega(x\in B_{a(n)}\rightarrow\exists m(x\in B_{b(m)})).$$
 Since the predicates ``$x\in B_{a(n)}$'' and ``$x\in B_{a(n)}$''
are open in $\mathcal{N}\times\mathcal{N}\times\omega$, the
predicate ``$\pi(a)\subseteq\pi(b)$'' is in
$\mathbf{\Pi}^1_1(\mathcal{N}\times\mathcal{N})$. Therefore
$E_\pi\in\mathbf{\Pi}^1_1(\mathcal{N})$.

From the previous paragraph it  follows that
$\pi^{-1}(\{A\})\in\mathbf{\Pi}^1_1(\mathcal{N})$ for each
$A\in\mathbf{\Sigma}^0_1(\mathcal{N})$, so in particular
$\pi^{-1}(\{\mathcal{N}\})\in\mathbf{\Pi}^1_1(\mathcal{N})$. For
the second assertion of the theorem it remains to show that any
$\mathbf{\Pi}^1_1(\mathcal{N})$-set is Wadge reducible to
$\pi^{-1}(\{\mathcal{N}\})$.

Recall that a  {\em tree in $\omega^\ast$} is any subset $T$ of
$\omega^\ast$ closed under prefixes. It is well known that the
closed subsets of $\mathcal{N}$ are precisely the sets
$[T]=\{x\in\mathcal{N}\mid\forall n(x[n]\in T)\}$ where $T$ ranges
though the trees in $\omega^\ast$, and that $[T]=\emptyset$ iff
$T$ is well founded, i.e. it contains no infinite ascending chain
$\tau_0\sqsubset\tau_1\sqsubset\cdots$. Furthermore,
$\mathcal{N}\setminus[T]=\bigcup\{\sigma\cdot\mathcal{N}\mid\sigma\in\partial
T\}$ where $\partial T$ is the set of minimal elements in
$(\omega^\ast\setminus T;\sqsubseteq)$. For any trees $T,S$, we
write $S\simeq T$ if there is an isomorphism $\varphi$ of
$(S;\sqsubseteq)$ onto $(T;\sqsubseteq)$ (note that we automatically have
$|\sigma|=|\varphi(\sigma)|$ for each $\sigma\in S$).

Let $W$ be the set of all $x\in\mathcal{N}$ such that the tree
$T_x=\{\tau\mid\exists n(\tau\sqsubseteq \sigma_{x(n)})\}$ is well founded.
It  is well known  (see e.g. Theorem 27.1 in \cite{ke94}) that $W$
is Wadge complete in $\mathbf{\Pi}^1_1(\mathcal{N})$, hence it
suffices to Wadge reduce $W$ to $\pi^{-1}(\{\mathcal{N}\})$.

It  is straightforward to define a continuous function $g$ on
$\mathcal{N}$ such that, for each $x\in\mathcal{N}$,
$\{\sigma_{g(x)(n)}\mid n<\omega\}=\partial S_x$ where $S_x$ is
some tree with $S_x\simeq T_x$. Then the continuous function $f$
on $\mathcal{N}$ defined by $f(x)(n)=g(x)(n)+1$, is a desired
Wadge reduction. Indeed, we have
 $$\pi f(x)=\bigcup_nB_{f(x)(n)}=\bigcup_n\sigma_{g(x)(n)}\cdot\mathcal{N}
 =\mathcal{N}\setminus[S_x],$$
 hence
\[x\in W\leftrightarrow[T_x]=\emptyset\leftrightarrow[S_x]=\emptyset
 \leftrightarrow\pi f(x)=\mathcal{N}.
\]

For the last assertion,  suppose that
$\mathbf{\Sigma}^0_1(\mathcal{N})$ is  countably based. By
Proposition 9 in \cite{br}, the equality relation on
$\mathbf{\Sigma}^0_1(\mathcal{N})$ is then $\mathbf{\Pi}^0_2$.
Since $\pi$ is continuous,
$E_\pi\in\mathbf{\Pi}^0_2(\mathcal{N})$, hence
$\pi^{-1}(\{\mathcal{N}\})\in\mathbf{\Pi}^0_2(\mathcal{N})$. This
contradicts to the second assertion of the theorem.
\qed

\begin{rems}\hfill
\begin{enumerate}[1.]
\item  As noted in Section \ref{hier}, for
quasi-Polish spaces the class ${\bf\Sigma}^1_{1}$ coincides with
the class of continuous images of Polish spaces. The last theorem
implies that this characterization cannot be extended to the
admissibly totally representable spaces because $\{\mathcal{N}\}$
is of course the image of a Polish spaces but it is not
${\bf\Sigma}^1_{1}$ (otherwise, we would get
$\pi^{-1}(\{\mathcal{N}\})\in\mathbf{\Sigma}^1_1(\mathcal{N})$
contradicting the third assertion of the theorem.)

\item  It may be shown  (as was noticed by M. de Brecht in a private
communication) that any sequential admissibly represented space
embeds into a sequential admissibly totally represented space
(namely into the space $\mathbf{\Sigma}^0_1(X)$ for a suitable
countably based space $X$). We hope that this result may be of use
for the development of DST for non-countably based spaces,
similarly to the use of the embeddability of all countably based
spaces into $P\omega$ for the development of DST for quasi-Polish
spaces \cite{br}.

\item Although the class  of sequential admissibly totally
represented spaces is rather rich (by the previous remark), it
does not form a cartesian closed category. This follows from
results in \cite{scs12} where, in particular, the smallest (in
some natural sense)  cartesian closed category of admissibly
represented spaces is identified.
\end{enumerate}
\end{rems}

\section{Semilattices of $\mathbf{\Sigma}^0_1$-Total Representations}\label{semilat}

A popular  field of numbering theory is the study of semilattices
of computable numberings of classes of computably enumerable sets.
This field is technically very complicated, even the
characterization of the simplest such semilattice --- the
semilattice of computably enumerable $m$-degrees --- is quite
hard. A long-standing open problem \cite{er77,er06} in this
field is to find invariants for the isomorphism relation on the
semilattices of computable numberings of finite classes of
computably enumerable sets.

In this section we discuss the topological analog of this field.
Again it turns out that the topological analog is much easier
(though non-trivial). We resolve the topological analog of a
problem related to the mentioned open problem of numbering theory.
This makes  use of some results mentioned in Section \ref{naming}.

Simplifying notation,  we denote
$\mathbf{\Sigma}^0_1(\mathcal{N})$ just by $\mathbf{\Sigma}^0_1$.
For $\mathcal{A}\subseteq\mathbf{\Sigma}^0_1$, let
$\mathcal{L}(\mathcal{A})$ (resp. $\mathcal{L}^\ast(\mathcal{A})$)
be the set of all $\mathbf{\Sigma}^0_1$-TRs of $\mathcal{A}$
(resp. the set of all $\mathbf{\Sigma}^0_1$-TRs
$\nu:\mathcal{N}\to\mathcal{A}$ of subsets of $\mathcal{A}$).
Let $L(\mathcal{A})$ (resp. $L^\ast(\mathcal{A})$) denote the
quotient-structure of the preorder
$(\mathcal{L}(\mathcal{A});\leq)$ (resp.
$(\mathcal{L}^\ast(\mathcal{A});\leq)$). Moreover, let
$L_\bot^\ast(\mathcal{A})$ be obtained by adjoining a new bottom
element $\bot$ to poset $L^\ast(\mathcal{A})$. We have the
following topological analog of a well known simple fact about
computable numberings.

 \begin{prop}\label{sem}\hfill
 \begin{enumerate}[\em(1)]
 \item $L(\mathcal{A})$ is an upper semilattice
 (in fact, a $\sigma$-semilattice).
 \item $L_\bot^\ast(\mathcal{A})$
is a distributive upper semilattice (in fact, a
$\sigma$-semilattice).
 \end{enumerate}
 \end{prop}

\proof Supremums  in both semilattices are obviously induced
by the operation $\oplus$. Distributivity means that if
$\xi\leq\mu\oplus\nu$ then $\xi\equiv\mu_1\oplus\nu_1$ for some
$\mu_1\leq\mu,\nu_1\leq\nu$ (the case of countable supremums is
considered similarly). If $\xi=\bot$, take $\mu=\nu=\bot$.
Otherwise, let $f$ be a continuous function on $\mathcal{N}$ that
reduces $\xi$ to $\mu\oplus\nu$. Let $A_0=\{a\in\mathcal{N}\mid
\exists n(f(a)(0)=2n)\}$ and $A_1=\{a\in\mathcal{N}\mid \exists
n(f(a)(0)=2n+1)\}$. Then $A_0,A_1$ are clopen and at least one of
them is non-empty. If $A_0=\emptyset$, take
$\mu_1=\bot,\nu_1=\xi$. If $A_1=\emptyset$, take
$\mu_1=\xi,\nu_1=\bot$. If both sets $A_0,A_1$ are non-empty, choose for
each $i<2$ a homeomorphism $f_i$ of $\mathcal{N}$ onto $A_i$ and
set $\mu_1=\xi\circ f_0$ and $\nu_1=\xi\circ f_1$. Then clearly
$\mu_1\leq\mu$ and $\nu_1\leq\nu$, so it remains to check that
$\xi\leq\mu_1\oplus\nu_1$. Define a continuous function $g$ on
$\mathcal{N}$ as follows: $g(x)=0\cdot f_0^{-1}(x)$ for $x\in
A_0$, and $g(x)=1\cdot f_1^{-1}(x)$ for $x\in A_1$. Then $g$
reduces $\xi$ to $\mu_1\oplus\nu_1$.
\qed

The semilattices  $L(\mathcal{A})$ and $L_\bot^\ast(\mathcal{A})$ might
be  quite complicated even for a countable set $\mathcal{A}$. But
if $\mathcal{A}$ is finite non-empty, the semilattices turn out to
be finite distributive lattices. The topological analog of the
mentioned problem from numbering theory is to find invariants for
$L(\mathcal{A})\simeq L(\mathcal{B})$ where $\simeq$ is the
isomorphism relation. This topological question seems to be much
easier than the mentioned problem (though we still do not know the
exact answer). E.g., from our results it follows that there is an
algorithm to answer the question $L(\mathcal{A})?\simeq
L(\mathcal{B})$ if the finite posets $(\mathcal{A};\subseteq)$ and
$(\mathcal{B};\subseteq)$ are given. The main result of this
section is the following theorem that gives very simple invariants
for the  relation $L^\ast(\mathcal{A})\simeq L^\ast(\mathcal{B})$.

 \begin{thm}\label{smain}
Let $\mathcal{A},\mathcal{B}$  be finite non-empty subsets of
$\mathbf{\Sigma}^0_1$. Then $L^\ast(\mathcal{A})\simeq
L^\ast(\mathcal{B})$ iff
$(\mathcal{A};\subseteq)\simeq(\mathcal{B};\subseteq)$.
  \end{thm}

This result  is a non-trivial corollary of some results in
\cite{he93,s04,s07a,ks07}. In the rest of this section we recall
some relevant information from those papers and deduce from them
the main result. First we recall necessary information from
\cite{s04} on $k$-labeled posets (see Section \ref{naming}).

For a finite poset $P\in\mathcal{P}$, let $rk(P)$  denote {\em the
rank of $P$}, i.e. the number of elements of the longest chain in
$P$. For any $1\leq i\leq rk(P)$,  let $P(i)=\{x\in P\mid
rk(\downarrow{x})=i\}$. Then $P(1),\ldots,P(rk(P))$ is a partition
of $P$ to ``levels''; note that $P(1)$ is the set of all minimal
elements of $P$. For any $x\in P$, let $suc(x)$ denote the set of
all immediate successors  of $x$ in $P$, i.e. $suc(x)=\{y\mid
x<y\wedge\neg\exists z(x<z<y)\}$. Note that $suc(x)=\emptyset$ iff
$x$ is maximal in $P$. The next result is Lemma 1.1 in \cite{s04}.

\begin{lem}\label{pos-for}
For any $P\in\mathcal{P}$  there exist $F=F(P)\in\mathcal{F}$ and a
monotone function $f$ from $F$ onto $P$ so that $rk(F)=rk(P)$,
$f$ establishes a bijection between $F(1)$ and $P(1)$, and for any
$x\in F$ $f$ establishes a bijection between $suc(x)$ and
$suc(f(x))$. The forest $F(P)$ is obtained by a natural bottom-up
unfolding of $P$.
 \end{lem}

Now we recall some information about {\em minimal $k$-forests} from
$\mathcal{F}_k$, i.e. $k$-forests not $h$-equivalent  to a
$k$-forest of lesser cardinality. The next fact is Lemma 1.3 in
\cite{s04}.

 \begin{lem}\label{min1}
Any two minimal $h$-equivalent $k$-forests  are isomorphic.
 \end{lem}

The next inductive characterization of  the minimal $k$-forests is
Theorem 1.4 in \cite{s04}.

\begin{lem}\label{min}\hfill
 \begin{enumerate}[\em(1)]
 \item  Any singleton $k$-forest is minimal.
 \item  A non-singleton $k$-tree $(T,c)$ is minimal iff $\forall x\in T(1)\forall
y\in T(2)(c(x)\not=c(y))$ and the $k$-forest $(T\setminus T(1),c)$ is minimal.
 \item  A proper $k$-forest is minimal iff all its $k$-trees are minimal and
pairwise incomparable under $\leq_h$.
 \end{enumerate}
  \end{lem}

\noindent For any finite non-empty set
$\mathcal{A}\subseteq\mathbf{\Sigma}^0_1$, let $k=|\mathcal{A}|$,
$\mathcal{A}=\{A_0,\ldots,A_{k-1}\}$, and $c(A_i)=i$ for each
$i<k$. Then we may think that $(\mathcal{A};\subseteq)$ is in
$\mathcal{P}$, the unfolding $F(\mathcal{A})$ of
$(\mathcal{A};\subseteq)$ is in $\mathcal{F}$,
$(\mathcal{A};\subseteq,c)$ is in $\mathcal{P}_k$, and
$(F(\mathcal{A});c\circ f)$ is in $\mathcal{F}_k$. The next lemma
follows from the previous one and the fact that the labeling
$c:\mathcal{A}\to k$ is bijective.

 \begin{lem}\label{unfold}
For any finite non-empty set
$\mathcal{A}\subseteq\mathbf{\Sigma}^0_1$,  the $k$-forest
$(F(\mathcal{A});c\circ f)$ is minimal.
 \end{lem}

There is a close relation of  $L^\ast(\mathcal{A})$ to the
difference hierarchy of $k$-partitions over the open sets. This
hierarchy developed in \cite{kw00,ko00,s04,s07a} extends from sets
to $k$-partitions the Hausdorff difference hierarchy over the open
sets. For any $P\in\mathcal{P}$, let $\mathbf{\Sigma}^0_1[P]$ be
the set of functions $\nu:\mathcal{N}\to P$ {\em defined by
$P$-families $\{A_p\}_{p\in P}$ of open sets}, i.e. there is a
family $\{A_p\}_{p\in P}$ of open sets such that $\nu(x)=p$ iff
$x\in A_p\setminus\bigcup\{A_q\mid p<q\}$,  for all $p\in
P,x\in\mathcal{N}$. For a $k$-poset $(P;d)\in\mathcal{P}_k$,
define the set $\mathbf{\Sigma}^0_1[P,d]$ of $k$-partitions of
$\mathcal{N}$ by
$\mathbf{\Sigma}^0_1[P,d]=\{d\circ\nu\mid\nu\in\mathbf{\Sigma}^0_1[P]\}$.

Items 1,2 of the following lemma  follow from Theorem 7.6 in
\cite{s07a}, item 3 follows from Theorem 3.1 in \cite{s04} (with a
heavy use of the $\omega$-reduction property of the open sets, see
Theorem \ref{uniform}), and item 4 follows from Lemma 5.1 in
\cite{s04}. For the definition of $\xi_G$ see Section
\ref{naming}.

\begin{lem}\label{dhpart}\hfill
 \begin{enumerate}[\em(1)]
 \item  For any $G\in\mathcal{F}$, $\mathbf{\Sigma}^0_1[G]=
\{\nu\in G^\mathcal{N}\mid \nu\leq \xi_G\}$, i.e. $\xi_G$ is a
complete element of $\mathbf{\Sigma}^0_1[G]$ with respect to
$\leq$.
 \item  For any $(G,d)\in\mathcal{F}_k$,
$\mathbf{\Sigma}^0_1[G,d]=\{\nu\in G^\mathcal{N}\mid \nu\leq
d\cdot\xi_G\}\subseteq (BC(\mathbf{\Sigma}^0_1))_k$.
 \item  For any $P\in\mathcal{P}$, $\mathbf{\Sigma}^0_1[P]=
 \{f\circ\nu\mid \nu\in \mathbf{\Sigma}^0_1[F(P)]\}$.
 \item For any finite non-empty set
$\mathcal{A}\subseteq\mathbf{\Sigma}^0_1$,
$\mathcal{L}^\ast(\mathcal{A})=\mathbf{\Sigma}^0_1[\mathcal{A},\subseteq]$.
 \end{enumerate}
  \end{lem}

Next we establish a close relationship of $L(\mathcal{A})$ and
$L^\ast(\mathcal{A})$ to some segments of the quotient-poset
$\mathbb{F}_k$ of the preorder $(\mathcal{F}_k;\leq_h)$. For
$a,b\in\mathbb{F}_k$, let $\downarrow{a}=\{x\mid x\leq_ha\}$  and
$[b,a]=\{x\mid b\leq_hx\leq_ha\}$. For any $i<k$, let $e_i$ be the
$h$-equivalence class of a singleton $k$-forest labeled by $i$, so
$\{e_0,\ldots,e_{k-1}\}$ is the enumeration without repetition of
the minimal elements of $\mathbb{F}_k$. Let
$e=e_0\sqcup\cdots\sqcup e_{k-1}$.

 \begin{prop}\label{isom}
For any finite non-empty set
$\mathcal{A}\subseteq\mathbf{\Sigma}^0_1$,
$L^\ast(\mathcal{A})\simeq\downarrow{a}$ and
$L(\mathcal{A})\simeq[e,a]$, where $a$ is the $h$-equivalence
class of $(F(\mathcal{A});\subseteq,c\circ f)$.
 \end{prop}

\proof An isomorphism between $\downarrow{a}$ and
$L^\ast(\mathcal{A})$ is  the restriction to $\downarrow{a}$ of
the function induced by the map $(G,d)\mapsto A\circ d\circ
\xi_G$. Indeed, if $(G,d)\leq_h(F(\mathcal{A});\subseteq,c\circ
f)$ then we subsequently deduce from Lemma \ref{dhpart} that
$\xi_G\in\mathbf{\Sigma}^0_1[G]$,
$d\circ\xi_G\in\mathbf{\Sigma}^0_1[G,d]$, $A\circ d\circ
\xi_G\in\mathbf{\Sigma}^0_1[\mathcal{A};
\subseteq]=\mathcal{L}^\ast(\mathcal{A})$. Since $A:k\mapsto
\mathcal{A}$ is a bijection, from Proposition \ref{kfor} we obtain
that $(G,d)\leq_h(G_1,d_1)$ is equivalent to $A\circ
d\circ\xi_G\leq A\circ d_1\circ\xi_{G_1}$.

For  the relation $L^\ast(\mathcal{A})\simeq\downarrow{a}$ we
still have to show that any $\nu\in\mathcal{L}^\ast(\mathcal{A})$
is equivalent to $A\circ d\circ\xi_G$ for some
$(G,d)\leq_h(F(\mathcal{A});\subseteq,c\circ f)$. By items 3 and 4
of Lemma \ref{dhpart}, $\nu=f\circ\mu$ for some
$\mu\in\mathbf{\Sigma}^0_1[F(\mathcal{A})]$, hence
$c\circ\nu=c\circ
f\circ\mu\in\mathbf{\Sigma}^0_1[F(\mathcal{A}),c\circ f]$.  By
items 1 and 2 of Lemma \ref{dhpart},
$c\circ\nu\in(BC(\mathbf{\Sigma}^0_1))_k$. By Proposition
\ref{kfor}, $c\circ\nu\equiv d\circ\xi_G$ for some
$(G,d)\in\mathcal{F}_k$, so it remains to show that
$(G,d)\leq_h(F(\mathcal{A});\subseteq,c\circ f)$. Suppose the
contrary, then
$d\circ\xi_G\not\in\mathbf{\Sigma}^0_1[F(\mathcal{A}),c]$ by  item
2 of Lemma \ref{dhpart}. By items 3 and 4 of Lemma \ref{dhpart},
$\nu\not\in\mathbf{\Sigma}^0_1[\mathcal{A}]=\mathcal{L}^\ast(\mathcal{A})$
which is a contradiction.

It remains to show that $L(\mathcal{A})\simeq[e,a]$. For any
$i<k$,  let $\mu_i=\lambda x.A_i$, then clearly
$\mu=\mu_0\oplus\cdots\oplus\mu_{k-1}$ is a smallest element in
$(\mathcal{L}(\mathcal{A});\leq)$. Moreover, the isomorphism above
sends $e$ to the equivalence class of $\mu$. Therefore the
restriction of that isomorphism to $[e,a]$ is a desired
isomorphism between $[e,a]$ and $L(\mathcal{A})$.
\qed

From the previous proposition  and Proposition \ref{sem} we
immediately obtain:

 \begin{cor}\label{dislat}\hfill
 \begin{enumerate}[\em(1)]
 \item For any finite non-empty set
$\mathcal{A}\subseteq\mathbf{\Sigma}^0_1$, $L^\ast(\mathcal{A})$
and $L(\mathcal{A})$ are finite distributive lattices.
 \item From given finite posets $(\mathcal{A};\subseteq)$
 and $(\mathcal{B};\subseteq)$ one can compute
whether $L^\ast(\mathcal{A})\simeq L^\ast(\mathcal{B})$ (or
$L(\mathcal{A})\simeq L(\mathcal{B})$).
  \end{enumerate}
 \end{cor}

\noindent We also need a result  on automorphisms of $\mathbb{F}_k$. Let
$Aut(\mathbb{F}_k)$ (resp. $Aut(k)$) denote the group of all
automorphisms of $\mathbb{F}_k$ (resp. of all permutations of
labels $0,\ldots,k-1$). For any $x\in\mathbb{F}_k$, let $M(x)$ be
the set of minimal elements of $\mathbb{F}_k$ below $x$ (this set
is in a bijective correspondence with the set of labels in some,
equivalently in any, $k$-forest in the $h$-equivalence class $x$).
Any permutation $p\in Aut(k)$ induces the automorphism
$(G,d)\mapsto(G,p\circ d)$ of $\mathbb{F}_k$ which is for
simplicity denoted by the same letter $p$.  We call elements
$x,y\in\mathbb{F}_k$  {\em automorphic} if $g(x)=y$ for some $g\in
Aut(\mathbb{F}_k)$.

 \begin{prop}\label{aut1}
For all  $x,y\in\mathbb{F}_k$, $\downarrow{x}\simeq\downarrow{y}$
 iff $x,y$ are automorphic.
  \end{prop}

\proof One  direction is obvious. Conversely,  it suffices
to show that for any isomorphism $h$ from $\downarrow{x}$ onto
$\downarrow{y}$ there is $p\in Aut(k)$ with $p(x)=h(x)$.  This is
checked by induction on the rank $rk(x)$ of $x$ in $\mathbb{F}_k$.
If $rk(x)=1$ then $x$ is minimal, hence $y$ is also minimal and
the assertion is obvious. The assertion is also easy in case
$|M(x)|=2$ because then $|M(y)|=2$ and the structure
$\mathbb{F}_2$ is almost well ordered and of rank $\omega$ (in
fact, it is isomorphic to the structure of finite levels of the
difference hierarchy of sets under inclusion). So assume
$|M(x)|\geq 3$ and consider two cases depending on whether $x$ is
join-irreducible in the distributive lattice $\mathbb{F}_k$
enriched by a bottom element.

If $x$ is not join-irreducible then $x=x_0\sqcup\cdots\sqcup x_n$
for some $n\geq 1$ and some join-irreducible pairwise incomparable
$x_0,\ldots,x_n<x$. Let  $h_i$ be the restriction of $h$ to
$\downarrow{x}_i$, then $h_i$ is an isomorphism $\downarrow{x}_i$
onto $\downarrow{y}_i$ for each $i\leq n$ where  $y_i=h(x_i)$. By
induction, there are $p_0,\ldots,p_n\in Aut(k)$ such that
$p_i(x_i)=h_i(x_i)=y_i$ for all $i\leq n$.  Then
$p_i(b)=h(b)=p_j(b)$ for all $i,j\leq n$ and $b\in M(x_i)\cap
M(x_j)$, hence there is $p\in Aut(k)$ such that $p(b)=p_i(b)$ for
all $i\leq n$ and $b\in M(x_i)$. Then $p(x_i)=p_i(x_i)=y_i$ for
all $i\leq n$, hence $p(x)=p(x_0)\sqcup\cdots\sqcup
p(x_n)=y_0\sqcup\cdots\sqcup y_n=y$.

Finally, let $x$ be  join-irreducible, hence $y$ is also
join-irreducible.  Let $x^\prime=\bigsqcup\{z\mid z<x\}$ and let $y^\prime$
be obtained similarly from $y$. Then $x^\prime<x$, $\forall
z<x(z\leq x^\prime)$ and similarly for $y$. By induction,
$p(x^\prime)=h(x^\prime)$ for some $p\in Aut(k)$. By
Lemma 5 in \cite{ks07}, the function $a\mapsto a^\prime$ on the
join-irreducible elements $a$ with $|M(a)|\geq 3$ is injective,
hence $y^\prime=h(x^\prime)$ and $p(x)=y$.
\qed


\proof[Proof of Theorem \ref{smain}]  It is easy to see that
$(\mathcal{A};\subseteq)\simeq(\mathcal{B};\subseteq)$ implies
$L^\ast(\mathcal{A})\simeq L^\ast(\mathcal{B})$. Conversely, let
$L^\ast(\mathcal{A})\simeq L^\ast(\mathcal{B})$. Then
$|\mathcal{A}|=k=|\mathcal{B}|$ because $|\mathcal{A}|$ and
$|\mathcal{B}|$ are the numbers of minimal elements in
$L^\ast(\mathcal{A})$ and $L^\ast(\mathcal{B})$, respectively. By
Proposition \ref{isom}, $\downarrow{a}\simeq\downarrow{b}$   where
$a$ and $b$ are the $h$-equivalence classes of the $k$-forests
$(F(\mathcal{A});\subseteq,c\circ f)$ and
$(F(\mathcal{B});\subseteq,c_1\circ f_1)$, respectively. By the
previous proposition, $p(a)=b$ for some $p\in Aut(k)$, i.e.
$(F(\mathcal{A});\subseteq, p\circ c\circ
f)\equiv_h(F(\mathcal{B});\subseteq,c_1\circ f_1)$. By  Lemmas
\ref {unfold} and \ref{min1}, the last $k$-posets are even
isomorphic via some isomorphism $\varphi:F(\mathcal{A})\to
F(\mathcal{B})$, so in particular $p\circ c\circ f=c_1\circ
f_1\circ\varphi$. Therefore, $A_i\mapsto B_{p(i)}$ is an
isomorphism of $(\mathcal{A};\subseteq)$ onto
$(\mathcal{B};\subseteq)$.
\qed

\section{Category of Total Representations}\label{categ}

Here we briefly discuss the category  $\mathcal{N}Set$ of TRs
(which is a topological version of the category of numbered sets
in numbering theory \cite{er73a,er75,er77}) and its relation to
the study of index sets and  $k$-partitions.

The category $\mathcal{N}Set$ is formed by arbitrary TRs as
objects and by the morphism between TRs defined as follows: a
{\em morphism} $f:\mu\to\nu$ of TRs $\mu$ and $\nu$ is a
function $f:\mu(\mathcal{N})\to\nu(\mathcal{N})$ such that
$f\circ\mu\leq\nu$ (in other words, $f\circ\mu=\nu\circ\hat{f}$
for some continuous function $\hat{f}$ on $\mathcal{N}$ called a
{\em realizer} of $f$ w.r.t. $\mu,\nu$).

Category $\mathcal{N}Set$ has some natural subcategories. E.g.,
relate to any equivalence relation $E$ on $\mathcal{N}$ the TR
$\kappa_E(x)=[x]_E=\{y\mid (x,y)\in E\}$ of the quotient-set
$\mathcal{N}/E$. Let $\mathcal{N}Eq$ be the full subcategory of
$\mathcal{N}Set$ with those $\kappa_E$ as the objects. The proof of
the next assertion is straightforward, so we give only a hint.

 \begin{prop}\label{cat}
The category $\mathcal{N}Set$ has countable products and coproducts
and is equivalent to the small category $\mathcal{N}Eq$.
\end{prop}

\proof[Proof Hint.]  For a sequence $\{\nu_n\}$ of TRs, let $P$
(resp. $Q$) be the Cartesian product (resp. the disjoint union) of
the sequence of sets $\{\nu_n(\mathcal{N})\}$. Then $P$ consist of
all sequences $(\nu_0(x_0),\nu_1(x_1),\ldots)$ where
$x_n\in\mathcal{N}$. The product $\nu$ of $\{\nu_n\}$ in
$\mathcal{N}Set$ is given by $\nu\langle
x_0,x_1,\ldots\rangle=(\nu_0(x_0),\nu_1(x_1),\ldots)$. The set $Q$
consists of all pairs $(n,\alpha_n(y))$ where
$n<\omega,y\in\mathcal{N}$. The coproduct $\mu$ of $\{\nu_n\}$ in
$\mathcal{N}Set$ is given by $\mu(n\cdot x)=(n,\nu_n(x))$.

The equivalence of categories $\mathcal{N}Set$ and $\mathcal{N}Eq$
is given by the inclusion functor $I:\mathcal{N}Eq\to\mathcal{N}Set$
and the kernel functor $K:\mathcal{N}Set\to\mathcal{N}Eq$ defined by
$K(\nu)=\kappa_{E_\nu}$ on objects (where $E_\nu=\{(x,y)\mid
\nu(x)=\nu(y)\}$) and by $K_f([x]_{E_\mu})=[f(x)]_{E_\nu}$ on
morphisms $f:\mu\to\nu$.
\qed

Let $\mathcal{N}Ad$ be the full subcategory of $\mathcal{N}Set$
formed by the admissible TRs $\alpha$ w.r.t. the final topology on
$\alpha(\mathcal{N})$. By a well known property of admissible
representations \cite{wei00}  (see also Theorem \ref{seq2}), the
morphisms of $\mathcal{N}Ad$ are precisely the continuous
functions. By Proposition \ref{adm},
$\alpha\mapsto\alpha(\mathcal{N})$ is a functor from
$\mathcal{N}Ad$ onto the category  of sequential topological
spaces having an admissible TR, with the continuous functions as
morphisms.

Note that, using other reducibilities from Section \ref{naming},
one can form some other categories of TRs, in particular the
categories $\mathcal{N}Set(\mathbf{\Delta}^0_\alpha)$ (resp.
$\mathcal{N}Set(\mathbf{\Delta}^1_1)$) which have the TRs as
objects and the functions realized by the
$\mathbf{\Delta}^0_\alpha$-functions (resp. by the
$\mathbf{\Delta}^1_1$-functions) on the names. We would like to
see some work on properties and applications of these categories.

We conclude this section by some remarks on the index sets and
$k$-partitions in topology. For  an arbitrary sequential
admissibly totally representable space $X$, we denote by
$\alpha_X$ an admissible TR of $X$. The topological complexity of
subsets $A$ of $X$ may be measured by the Wadge degree of its
index set $\alpha^{-1}_X(A)$: the structure of Wadge degrees
guarantees that this complexity is essentially an ordinal. A similar
situation is well known in computability theory but there, because
of the complexity of the structure of $m$-degrees, the complexity
of a set is measured not by the $m$-degree of its index set (which
is the computable analog of the Wadge degree) but rather by the
position of the index set in a suitable hierarchy. Note that in
the study of index sets we again see the advantage of TRs against
representations because the Wadge degree of an index set in a partial
representation depends not only on the set $A$ but also on the
domain of the representation.

Note that, in contrast with computability theory, the topological
complexity of $A\subseteq X$ may be in principle measured
``directly'' by the Wadge degree of $A$ in the structure
$(P(X);\leq^X_W)$. But here we get the obstacle that for many
spaces $X$ the structure of Wadge degrees of subsets of $X$ is
complicated (in particular, this applies to the space of reals
\cite{he96}), so we again may have no convenient scale to measure
the topological complexity. For these reasons the index set
approach is often more useful. Note that
$A\mapsto\alpha^{-1}_X(A)$ is a homomorphism from
$(P(X);\leq^X_W)$ into $(P(\mathcal{N});\leq_W)$.

The mentioned approach to topological complexity may be in a
straightforward way extended to the study of topological
complexity of $k$-partitions of $X$ (and even of more complex
functions on spaces). Relate to any $k$-partition $A:X\to k$ the
$k$-partition $\alpha_X\circ A$ of $\mathcal{N}$. We call
$\alpha_X\circ A$ the  {\em index $k$-partition} of $A$ (cf.
\cite{s05}) because for $k=2$ the index $k$-partitions essentially
coincide with the index sets. The topological complexity of $A$ is
measured by the equivalence class of $\alpha_X\circ A$ in the
quotient-structure of $(k^\mathcal{N};\leq)$, see Section
\ref{naming}. This suggests a way to measure the topological
complexity of $k$-partitions, and to compare the complexity of
$k$-partitions of different spaces. E.g., for $k$-partitions
$A:X\to k$ and $B:Y\to k$ of quasi-Polish spaces $X,Y$ we say that
$A$ is {\em explicitly reducible} (resp. {\em implicitly
reducible}) to $B$ if $A=B\circ f$ for a continuous function
$f:X\to Y$ (resp. if $\alpha_X\circ A\leq\alpha_Y\circ B$). Note
that if $A$ is explicitly reducible to $B$ then it is also
implicitly reducible. In \cite{s82} similar concepts (called there
generalized index sets and reducibility by morphisms) were
introduced and studied in the context of computability theory.

We give an example from \cite{he96} relevant to CA which illustrates
the above notions. Let $\mathbb{C}$ be the space of complex
numbers and, for each $n\geq 1$, let $\mathbb{P}_n$ be the set of
polynomials $p=a_0+a_1z+\cdots+a_{n-1}z^{n-1}+z^n$ with complex
coefficients; $\mathbb{P}_n$ may be considered as a space
homeomorphic to $\mathbb{C}^n$. Define functions
$c:\mathbb{C}^n\to\{1,\ldots,n\}$ and
$r:\mathbb{P}_n\to\{1,\ldots,n\}$ as follows:
$c(z_0,\ldots,z_{n-1})$ is the cardinality of the set
$\{z_0,\ldots,z_{n-1}\}$, and $r(p)$ is the cardinality of the set
of complex roots of $p$. Then $c$ is explicitly reducible to $r$
(a reduction is given by the Vieta map
$(z_0,\ldots,z_{n-1})\mapsto(z-z_0)\cdots(z-z_{n-1})$). We do not
know whether $r$ is explicitly reducible to $c$ but certainly $r$
is implicitly reducible to $c$ (via a function that computes from
an $\alpha_{\mathbb{P}_n}$-name of a polynomial some
$\alpha_{\mathbb{C}^n}$-name of a vector of all its roots).
Therefore, $\alpha_{\mathbb{C}^n}\circ
c\equiv\alpha_{\mathbb{P}_n}\circ r$, hence the complexity of
these problems is measured by the same element of the
quotient-structure of $(k^\mathcal{N};\leq)$, in fact of
$((BC(\mathbf{\Sigma}^0_1))_k;\leq)$. In a sense, this shows that
functions $c$ and $r$ have the same topological complexity (called
discontinuity degree in \cite{he93}). By the result of P. Hertling
in \cite{he93} mentioned in Section \ref{naming}, this complexity
is characterized by a finite $k$-labeled forest, and this forest
(in fact, a linear order) is computed in \cite{he96}. Note that
TRs $\alpha_{\mathbb{C}^n}$ and $\alpha_{\mathbb{P}_n}$ may be
chosen so that the equivalence $\alpha_{\mathbb{C}^n}\circ
c\equiv\alpha_{\mathbb{P}_n}\circ r$ holds even effectively, i.e.
there are computable reductions in both directions.

\section{Reducibilities of Equivalence Relations}\label{equiv}

A popular topic in DST is the study of some reducibilities on
equivalence relations on the Baire space (see e.g.
\cite{ka08,gao09} for surveys). Here we note that these
reducibilities fit well to our framework and answer a natural
question for some of the corresponding degree structures.

The most popular reducibilities on equivalence relations are
defined as follows. For equivalence relations $E,F$ on
$\mathcal{N}$, $E$ is {\em continuously} (resp. {\em Borel})
reducible to $F$, in symbols $E\leq_cF$ (resp. $E\leq_BF$) if
there is a continuous (resp. a Borel) function $f$ on
$\mathcal{N}$ such that for all  $x,y\in\mathcal{N}$, $E(x,y)$ is
equivalent to $F(f(x),f(y))$. Note that these reducibilities are
closely related to (in fact, are strengthenings of) the
corresponding explicit reducibilities from the previous section.

The structures $(ER(\mathcal{N});\leq_c)$ and
$(ER(\mathcal{N});\leq_B)$ where $ER(\mathcal{N})$ is the set of
all equivalence relations on $\mathcal{N}$, and especially their
substructures on the set of Borel equivalence relations, were
intensively studied in DST. In particular, it was shown that both
structures are rather rich. But, to my knowledge, no result about
the complexity of first-order theories of these structures and
their natural substructures was established so far. Such results
are desirable, as the history of degree structures in
computability theory demonstrates.

Below we show that most of the natural substructures of the first
structure have undecidable first-order theories (unfortunately, our
methods do not apply to the second structure, so for the Borel
reducibility the question remains open). We concentrate first on
the initial segment $(ER_k;\leq_c)$ of $(ER(\mathcal{N});\leq_c)$
formed by the set $ER_k$ of equivalence relations which have at
most $k$ equivalence classes. We relate this substructure to the
structure $(k^\mathcal{N};\leq_1^\prime)$ where $\leq_1^\prime$ is
the following slight modification of the reducibility $\leq_1$ in
Section \ref{naming}: $\mu\leq_1^\prime\nu$ iff
$\mu=\varphi\circ\nu\circ f$ for some continuous function $f$ on
$\mathcal{N}$ and for some permutation $\varphi$ of
$\{0,\ldots,k-1\}$.

 \begin{prop}\label{er1}
For any $2\leq k<\omega$, the function $\nu\mapsto E_\nu$ induces
an isomorphism between the quotient-structures of
$(k^\mathcal{N};\leq_1^\prime)$ and $(ER_k;\leq_c)$.
\end{prop}

\proof First we check that $\mu\leq_1^\prime\nu$ iff
$E_\mu\leq_cE_\nu$ via $f$. Let $\mu\leq_1^\prime\nu$, so
$\mu=\varphi\circ\nu\circ f$ for some continuous function $f$ on
$\mathcal{N}$ and for some permutation $\varphi$ of
$\{0,\ldots,k-1\}$. Then $E_\mu\leq_cE_\nu$ via $f$.

Conversely, let $E_\mu\leq_cE_\nu$ via $f$. Define the function
$\psi:\mu(\mathcal{N})\to k$ by $\psi(\mu(x))=\nu f(x)$. Since
$\mu(x)=\mu(y)$ implies $\nu f(x)=\nu f(y)$, $\psi$ is correctly
defined. Since $\mu(x)\not=\mu(y)$ implies $\nu f(x)\not=\nu
f(y)$, $\psi$ is injective. Let $\varphi$ be a permutation of $k$
so that $\varphi\psi(i)=i$ for each $i\in\mu(\mathcal{N})$. Then
$\varphi\nu f(x)=\varphi\psi\mu(c)=\mu(x)$, hence
$\mu\leq_1^\prime\nu$.

To complete the proof, it suffices to show that for any $E\in
ER_k$ there is $\nu\in k^\mathcal{N}$ with $E=E_\nu$. Let
$(E_0,\ldots,E_i)$ be an enumeration without repetition of the
equivalence classes of $E$. Define $\nu:\mathcal{N}\to\{0,\ldots,i\}$ by
$\nu(x)=j\leftrightarrow x\in E_j$, for all $j\leq i$ and
$x\in\mathcal{N}$. Then $E=E_\nu$.
\qed

 \begin{thm}\label{er2}
Let $k\geq3$ and let $A$ be any initial segment of
$(ER(\mathcal{N});\leq_c)$ that contains all relations in
$ER_k\cap BC(\mathbf{\Sigma}^0_1(\mathcal{N}))$. Then the first-order theory
of the quotient-structure of $(A;\leq_c)$ is undecidable.
 \end{thm}

\proof Let $B=\{\nu\in k^\mathcal{N}\mid E_\nu\in A\}$. By
the previous proposition it suffices to show that the first-order
theory of the quotient-structure of $(B;\leq_1^\prime)$ is
undecidable. By Theorem 2 in \cite{ksz10}, the first-order theory
of the quotient-structure of $(B;\leq_1)$ is undecidable. An
inspection of that proof shows that it also works for the relation
$\leq_1^\prime$.
\qed

\section{Conclusion}\label{con}

We hope that this paper demonstrates that total representations
deserve  special attention because they are sufficient to
represent many spaces of interest, appear naturally as the
principal TRs of levels of the popular hierarchies, simplify
and uniform presentation of some topics, suggest new open
questions and make a much better analogy with the numbering theory
than the partial representations. At the same time, there are
several important topics (in particular, complexity in analysis,
functionals of finite type or the study of rich enough cartesian
closed categories of spaces) where partial representations are
really inevitable.

{\bf Acknowledgement.} I am grateful to Matthew de Brecht for
providing a preliminary copy of his paper \cite{br}. I  also thank
Vasco Brattka, Peter Hertling and Matthias Schr\"oder for
stimulating discussions, and to the anonymous referees for many
useful suggestions and bibliographical hints.

\end{document}